\documentclass[journal=jctcce,manuscript=article]{achemso}

\usepackage[version=3]{mhchem} 
\usepackage{listings}
\usepackage{widetext}
\usepackage{tikz}
\usetikzlibrary{shapes.geometric, arrows}
\usetikzlibrary{shapes.misc}
\usetikzlibrary{fit,backgrounds}
\usepackage{enumitem}

 \usepackage[normalem]{ulem} 

\tikzstyle{tbb_process} = [draw,dashed,black,rounded corners,fill=yellow!50]
\tikzstyle{process} = [rectangle, minimum width=3cm, minimum height=1cm, text centered, draw=black, fill=orange!30]
\tikzstyle{decision} = [chamfered rectangle, minimum width=3cm, minimum height=1cm, text centered, draw=black, fill=green!30]
\tikzstyle{comm} = [rectangle, rounded corners, minimum width=3cm, minimum height=1cm,text centered, draw=black, fill=red!30]
\tikzstyle{loop} = [rounded rectangle, minimum width=3cm, minimum height=1cm, text centered, draw=black, fill=blue!30]
\tikzstyle{io} = [text centered]
\tikzstyle{arrow} = [thick,->,>=stealth]
\tikzstyle{carrow} = [very thick,->,>=stealth, red]
\tikzstyle{sarrow} = [very thick, double, red]

\graphicspath{{Figs/}}

\usepackage{xcolor}
\newcommand{\ToDo}[1]{{\color{black} #1}}

\newcommand{\TRB}[1]{\,\textrm{Tr$_{\rm B}$}\left( #1 \right)\,}
\newcommand{\MXEL}[3]{\left< #1 | #2 | #3 \right>}
\newcommand{\PRJ}[2]{| #1 \rangle \langle #2 |}
\newcommand{\KET}[1]{\left.| #1 \right>}

\newcommand{\EXP}[1]{{\rm exp}\left[#1\right]}

\author{Roman Ovcharenko}
\email{roman.ovcharenko@cup.lmu.de}
\author{Benjamin P. Fingerhut}
\email{benjamin.fingerhut@cup.lmu.de}
\affiliation[LMU]{Department of Chemistry and Centre for NanoScience, Ludwig-Maximilians-Universität München, 81377 München, Germany.}

\date{\today}

\title[MACGIC-iQUAPI]{Scalable Distributed Memory Implementation of the Quasi-Adiabatic Propagator Path Integral}
\abbreviations{IR,NMR,UV}

\begin{document}



%
%



\begin{abstract}
The accurate simulation of dissipative quantum dynamics subject to a non-Markovian environment poses persistent numerical challenges, in particular for structured environments where sharp mode resonances induce long-time system bath correlations.  
We present a scalable distributed memory implementation of the 
Mask Assisted Coarse Graining of Influence Coefficients (MACGIC) - Quasi-Adiabatic Propagator Path Integral (-QUAPI) method 
that exploits the memory resources of multiple compute nodes and mitigates  the memory bottleneck of the method via a new pre-merging algorithm while preserving numerical accuracy.
The distributed memory implementation spreads the paths over the computing nodes by means of the MPI protocoll and efficient high level path management is achieved via an implementation based on hash maps.
The efficiency of the new implementation is demonstrated in large-scale dissipative quantum dynamics simulations that account for the coupling to a structured non-Markovian environment containing a sharp resonance, a setup for which convergence properties are investigated in depth.
Broad applicability and the non-perturbative nature of the simulation method is illustrated
via the tuning of the  mode resonance frequency of  the structured environment
with respect to  the system frequency. 
The simulations reveal a splitting of resonances due to strong system-environment interaction and the emergence of sidebands due to 
multi-excitations of the bosonic mode that are not accounted for in perturbative approaches. 
The simulations demonstrate the versatility of the new MACGIC-QUAPI method in the presence of strong non-Markovian system bath correlations.
\end{abstract}

\section{Introduction}

The time evolution of a quantum system is decisively affected by the interaction with the environment through which, regardless of the precise nature of the environment, phenomena like energy relaxation and dephasing arise.\cite{b:Weiss_08, b:Petruccione_02}.
For example in  condensed matter, phonons are responsible for the dissipation of energy~\cite{b:Kuehn_11}. In quantum computing, undesirable decoherence of superconducting qubits arises from the inevitable coupling to the electromagnetic environment~\cite{a:Shnirman_01, a:Wal_00, a:Mooij_03}. Similarly, the precise nature of energy or charge transfer dynamics 
can only be rationalized via the interaction with the environment and theoretical descriptions have to account for the nuclear degrees of freedom of the solvent or a biological environment \cite{Jortner:AdvChem:1999,Tamura:JPCC:2011}.


The finite memory time of the environment induces non-Markovian  behavior by  
 imposing  correlations  between the system  and the environment 
 that affect decoherence and transfer rates.\cite{Vega:RevModPhys:2017} Under conditions where 
 intersite  and  system-environment interactions are of  similar magnitude  \cite{Ishizaki:JCP:2009} or the dynamics of the environment is slow \cite{Nalbach:JCP:2010},
 the Born-Markov approximation is no longer valid and  the finite  bath  memory  time imposes strong correlations that persist on a  time scales that is comparable to the bare system dynamics.
%
Particular 
challenges arise  when the quantum system interacts with a structured environment, e.g, the discrete modes of the nuclear degrees of freedom or the dispersive readout modes of superconducting qubits.
Such structured environments can potentially induce  long-time memory for the interaction between the system and the environment and the underlying non-Markovian memory effects are difficult to capture by methods rooted in perturbation theory. The accurate simulation of the real-time non-equilibrium  quantum dynamics subject to such non-Markovian dissipation is a persistent numerical challenge.\cite{Rosenbach:NJP:2016,Iles-Smith:JCP:2016,Gibben:PhysRevResearch:2020}

The path integral formulation\cite{a:Feynmann_48} of quantum mechanics provides a non-perturbative description of the time evolution of the reduced density matrix that is particularly appealing for harmonic baths where the environmental degrees of freedom can be integrated analytically  
in the form of the Feynmann–Vernon inﬂuence functional.\cite{a:Vernon_63}
%
For example, the Hierarchy of Equation of Motions (HEOM)~\cite{a:Kubo_89, a:Tanimura_90} method maps the effect of the non-Markovian system–bath interaction onto the hierarchical elements of the reduced density matrix.\cite{Tanimura:JCP:2020}
Alternatively,  by assuming a finite time span of non-Markovian memory, an augmented reduced density matrix tensor
is iteratively constructed in the   quasi-adiabatic propagator path integral (QUAPI) method~\cite{a:Makarov_th_95, a:Makarov_res_95}.
While in principle being applicable to arbitrary functional forms of the spectral density, 
the size of the augmented reduced density matrix tensor scales exponentially with the system-bath correlation time, and  the applicability of the QUAPI method is thus generally limited to cases where system-bath correlations decay very rapidly and the spectral density function does not include  discrete underdamped modes of the environment.

A number of approaches recently addressed longtime system-bath correlations that arise from a '\emph{sluggish}' environment with long non-Markovian memory.\cite{a:Lambert_12}
The time-evolving matrix product operators (TEMPO)~\cite{Strathearn:NatCom:2018} method 
adopts matrix product states~\cite{Schollwock:AnnPhys:2011,Orus:AnnPhys:2014} for the 
representation of the augmented density matrix tensor. 
The small matrix decomposition of the path integral (SMatPI) approach of Makri~\cite{a:Makri_20}  exploits a recursive decomposition of the augmented density matrix tensor and avoids 
the exponentially increasing memory requirements of QUAPI via the recursive spread of the entangled influence functional terms over longer time intervals. 
Recently, we have introduced the mask assisted coarse graining of influence coefficients  (MACGIC-QUAPI)~\cite{a:Fingerhut_17} that exploits  
physical properties of the bath correlation function, i.e., a fast decay at early times and a slower temporal variation at later times. This behavior facilitates an intermediate coarse grained representation of the influence functional and allows for a decoupling of the total system bad correlation time from the number of considered Feynman paths. 
MACGIC-QUAPI simulations provide numerical access to long system-bath correlation times and allow for accurate simulations of relatively complex systems, like energy equilibration in the FMO trimer complex~\cite{a:Fingerhut_17} and  the  coupled excitation energy and charge transfer dynamics in a model of the bacterial reaction center.\cite{a:Fingerhut_19} 

Besides Refs.\citenum{a:Makri_97_II} and~\citenum{a:Sato_19}, computational implementation aspects of the QUAPI method are largely unaddressed.  
Here,  we present a  scalable distributed memory implementation in conjunction with a novel pre-merging algorithm of the MACGIC-QUAPI method. 
Distributed memory parallelization allows to exploit physical memory resources of different compute nodes while the inherent non-local structure of the MACGIC-QUAPI algorithm requires careful synchronization, ordering and merging of paths.
The optimized 
pre-merging algorithm reduces the memory bottleneck of the method and further facilitates the distributed memory implementation.
Numerical  efficiency of the implementation is demonstrated in large scale dissipative quantum dynamics simulations of a quantum system subject to strong interaction with a structured non-Markovian environment. 
By varying the frequency  of the sharp environment resonance with respect to the characteristic frequency of the quantum system, 
 profound non-Markovian effect in the system dynamics  are revealed that demonstrate the limits of perturbative approaches. 

The remainder of the paper is organized as follows: 
 Section~\ref{sec:method} gives a description of the QUAPI formalism and summarizes 
 the MACGIC-QUAPI method.
Section~\ref{sec:impl}   presents the new path pre-merging algorithm and describes the parallel distributed memory implementation.
Numerical accuracy of the path pre-merging algorithm is demonstrated in Section~\ref{Sec:Premerge}  and the performance of the parallel distributed memory implementation is analyzed in Section~\ref{sec:DistrMem}. 
The developed program is finally applied to investigate the impact of a non-Markovian environment with 
sharp resonance frequency on the dynamics of a two-level quantum system (Sec.~\ref{sec:struc_env}).
%

\section{Quasi-Adiabatic Propagator Path Integral (QUAPI) Method}
\label{sec:method}

We consider a generic system Hamiltonian linearly coupled to a harmonic bath~\cite{b:Weiss_08, b:Kuehn_11}:
%
%
\begin{equation}
\begin{aligned}
&\hat H = \hat H_S \otimes \mathbf{\hat I_B} + \hat I_S \otimes \mathbf{\hat H_{B}} + \mathbf{\hat H_{SB}} 
\\
&\hat H_S = \sum_{i j} h_{i j} \, \PRJ{\tilde q_i}{\tilde q_j}
&\mathbf{\hat H_B} = \sum_j \left( \frac{\mathbf{\hat P_j}^2}{2 m_j} + \frac{1}{2} m_j w_j^2 \mathbf{\hat Q_j}^2 \right) 
\\
&\mathbf{\hat H_{SB}} = - \sum_{i} q_{i} \PRJ{\tilde q_i}{\tilde q_i} \otimes \sum_{j} c_j \mathbf{\hat Q_j}.
\end{aligned}
\label{eq:tot_ham}
\end{equation}
Here, 
 we adopt the bra-ket notation for the system  while bath operators  are indicated with bold font.
 $\hat H_S$ denotes the system 
 Hamiltonian, commonly employed in discrete variable representation (DVR)~\cite{a:Makri_93} $\KET{\tilde q_i}$.  The bath Hamiltonian  $\mathbf{\hat{H}_B}$ describes a  continuous set of harmonic oscillators with momentum and position operators $\mathbf{\hat P_j}$ and $\mathbf{\hat Q_j}$, respectively. 
 $\mathbf{\hat H_{SB}}$ describes the bilinear system-bath coupling via the diagonal elements of $\hat H_S$, where  $c_j$  characterizes the  coupling strength between system and environment
 and collective  system  coordinate $q$ of the quantum system.

The relevant information concerning the time evolution of the system subject to the interaction with the environment is contained in the reduced density matrix $\hat \rho(t)$ that is obtained upon tracing over the bath degrees of freedom:
\begin{equation}
\hat \rho(t) = \TRB{e^{-i \mathbf{\hat H} t / \hbar} \cdot \boldsymbol{\hat \rho}(0) \cdot e^{i \mathbf{\hat H} t / \hbar}}.
\label{eq:red_rho}
\end{equation}
In the following we consider factorized initial conditions $\boldsymbol{\hat \rho}(0) = \hat \rho(0) \otimes \boldsymbol{\hat \rho}(0)$ with the bath initially in thermal equilibrium, i.e.,
 $\boldsymbol{\hat \rho}(t=0) = e^{-\mathbf{\hat H_B} / k_B T} / \TRB{e^{-\mathbf{\hat H_B} / k_B T}}$.~\cite{a:Leggett_83, a:Zwerger_87}

All information about the environment is contained in the  spectral density function 
%
\begin{equation}
J(\omega) = \frac{\pi}{2} \sum_j \frac{c_j^2}{m_j \omega_j} \delta(\omega-\omega_j),
\label{eq:SD}
\end{equation}
with the reorganization energy 
\begin{equation}
\lambda = \int_0^\infty \frac{J(\omega)}{\omega} d\omega 
\end{equation}
providing a measure of the overall interaction strength of system and environment.

The quasi-adiabatic path integral (QUAPI) method 
relies on a repartitioning of  the total Hamiltonian (eq.~\ref{eq:tot_ham}):\cite{a:Makri_92}
 \begin{widetext}
\begin{equation}
\begin{aligned}
&\hat H = \hat H_S^a \otimes \mathbf{\hat I_B} + \mathbf{\hat H_{B}^a}(q)
\\
&\hat H_S^a = \sum_{i j} h_{i j} \, \PRJ{\tilde q_i}{\tilde q_j} - \sum_j \frac{c_j^2}{2m_j w_j^2} \, \sum_i q_i^2 \, \PRJ{\tilde q_i}{\tilde q_i} 
\\
&\mathbf{\hat H_B^a}(q) = \sum_j \left( \hat I_S \otimes \frac{\mathbf{\hat P_j}^2}{2 m_j} + \frac{1}{2} m_j w_j^2 \left( \hat I_S \otimes \mathbf{\hat Q_j} - \frac{c_j}{m_j w_j^2} \, \sum_i q_i \, \PRJ{\tilde q_i}{\tilde q_i} \otimes \mathbf{\hat I_B} \right)^2 \right)
\end{aligned}
\label{eq:tot_new}
\end{equation}
\end{widetext}
where $\hat H_S^a$ is a renormalized system Hamiltonian along an adiabatic path and $\mathbf{\hat H_B^a}(q) $ 
is the adiabatically displaced bath Hamiltonian. 
Applying a symmetric Trotter splitting to Eq.~(\ref{eq:red_rho}), the general QUAPI expression for the reduced density matrix is obtained~\cite{a:Makri_95}:
\begin{align}
\MXEL{\tilde q_N^+}{\hat \rho(t)}{\tilde q_N^-} = \sum_{q_0^\pm} \MXEL{\tilde q_0^+}{\hat \rho(0)}{\tilde q_0^-}  \times \sum_{\{q_j^\pm \}} S(q_0^\pm, \{q_j^\pm\}, q_N^\pm, t)  \times I(q_0^\pm, \{q_j^\pm\}, q_N^\pm, t).
\label{eq:quapi_red_rho}
\end{align}
Here the  bare system propagator $S(q_0^\pm, \{q_j^\pm\}, q_N^\pm, t) $ for finite time $t = N \Delta t$ takes the form
\begin{equation}
S(q_0^\pm, \{q_j^\pm\}, q_N^\pm, t) =  \prod_{j=1}^N S(q_j^\pm, q_{j-1}^\pm).
\label{eq:SysP0}
\end{equation}
with the Trotter propagation time  step $\Delta t$ and the short-term system propagator 
\begin{align}
S(q_j^\pm, q_{j-1}^\pm) = \MXEL{\tilde q_j^+}{e^{-i \hat H_S^a \Delta t / \hbar}}{\tilde q_{j-1}^+}  \times  \MXEL{\tilde q_{j-1}^-}{e^{i \hat H_S^a \Delta t / \hbar}}{\tilde q_{j}^-} ,
\label{eq:SysP}
\end{align}
describing the time evolution of the isolated system along an adiabatic path, neglecting effects of the  bath.
In eq.~\ref{eq:SysP},
$q_{j-1}^\pm$ and $q_j^\pm$ denote forward ($+$) and backward ($-$) propagation coordinates at time $t$ and $t+\Delta t$, respectively.
The sum $\sum_{\{q_j^\pm \}}$ in eq.~\ref{eq:quapi_red_rho} runs over all possible quantum mechanical paths between the fixed initial forward-backward points $q_0^\pm$ and the final points $q_N^\pm$.
%

Non-Markovian effects on the dynamics are captured by the action of the  discretized Feynman-Vernon influence functional~\cite{a:Vernon_63, a:Makri_95}
%
\begin{align}
I(q_0^\pm, \{q_j^\pm\}, q_N^\pm, t)  = \EXP{-\frac{1}{\hbar} \sum_{i = 0}^{N} \sum_{j = 0}^{i} \left( q_i^+ - q_i^- \right) \left( \eta_{i j} \, q_j^+ - \eta_{i j}^* \, q_j^- \right) } ,
\label{eq:inf_fun}
\end{align}
    where $\{\eta_{i j}\}$ are the discretized analog of the bath response function~\cite{a:Makri_97} that are determined by the bath spectral density (eq.~\ref{eq:SD}) (explicit expressions are given Refs.~\citenum{a:Makarov_th_95,a:Fingerhut_17}).
%
%
%
The influence functional introduces 
non-Markovian effects due to the time non-local correlations between system coordinates.

For a sufficiently small Trotter time step $\Delta t$, eq.~\ref{eq:inf_fun} provides an exact description of the time evolution of a quantum system bilinearly coupled to a bath, without restriction to a particular form  of the spectral density (eq.~\ref{eq:SD}). 
Nevertheless, numerical applications are limited by the exponentially growing  
number of paths with propagation time, imposing  
a $O(M^{2N})$ numerical scaling, with $M$ being the number of system states and $N$ the number of time steps (eqs. \ref{eq:SysP0},\ref{eq:inf_fun}). 
Starting from eq.~\ref{eq:inf_fun}, the
 QUAPI 
method relies on a finite time span of the non-Markovian memory to iteratively construct the propagator tensor.\cite{a:Makri_94, a:Makarov_th_95, a:Makarov_res_95}
During the memory time, the magnitude of correlation between two time points $i$ and $j$, reflected in $\eta_{i j}$, drops significantly for the large separation $d = i - j$, 
and becomes negligible for some threshold value $\Delta k_{\rm max} \ll N$. 
The assumption of a finite memory time $\tau = \Delta k_{\rm max} \Delta t$ is physically well motivated by the broad environmental spectra in condensed phase and formally represents the only limitation of the method.
%


\subsection{Mask Assisted Coarse Graining of Influence Coefficients \\ (MACGIC-QUAPI)} 
\label{sec:appx}

The  exponential scaling of the QUAPI method for increasing memory time $\tau$  was addressed by introducing a coarse grained intermediate representation of the influence functional.\cite{a:Fingerhut_17}
The idea behind the method is based on generic properties of typical bath response functions, i.e, the observation that influence coefficients $\eta_{i j}$ show a faster decay for small time lags $d = i - j$ 
than for large ones.
 It is thus possible to construct an efficient numerical representation that relies on a non-uniform temporal grid where a denser temporal spacing is employed for small time lags $d$ and a sparser  temporal spacing is employed for large time lags $d$. 
 Since the original bath memory grid is uniform, an auxiliary non-uniform coarse grained time grid with 
 $\Delta k_{\rm eff} \ll \Delta k_{\rm max}$ 
 elements is introduced that is selected via a mask from the uniform grid (mask assisted coarse graining of influence coefficients, MACGIC-QUAPI).
 
The  auxiliary grid is employed  for a reduction of number of paths used for propagation via merging~\cite{a:Makri_12,a:Fingerhut_17} while the original uniform time grid is employed for propagation of the reduced density matrix. 
Since the scaling of the MACGIC-QUAPI method is determined by the mask size 
$\Delta k_{\rm eff}$, $O(M^{2\Delta k_{\rm eff}})$, 
that is typically substantially smaller than 
$\Delta k_{\rm max}$, the algorithm allows to decouple the numerical effort from memory time $\tau$. The algorithm thus provides access to longtime memory effects of a ’\emph{sluggish}’ environment beyond the capabilities of the standard QUAPI method and convergence to the latter result is retained for $\Delta k_{\rm eff} \rightarrow \Delta k_{\rm max}$. 
Details of the mask optimization procedure are given in Ref.~\citenum{a:Fingerhut_17} and applicability to non-monotonous decaying bath response functions was demonstrated in Ref.~\citenum{a:Fingerhut_19}.

While the underlying idea of the MACGIC-QUAPI method is appealingly transparent, its numerical implementation poses some challenges. 
Scheme~\ref{sch:macgic} summarizes the working principle of the MACGIC-QUAPI algorithm and the employed subroutines.
%
\begin{scheme*}[t!]
\includegraphics[width=0.92\textwidth]{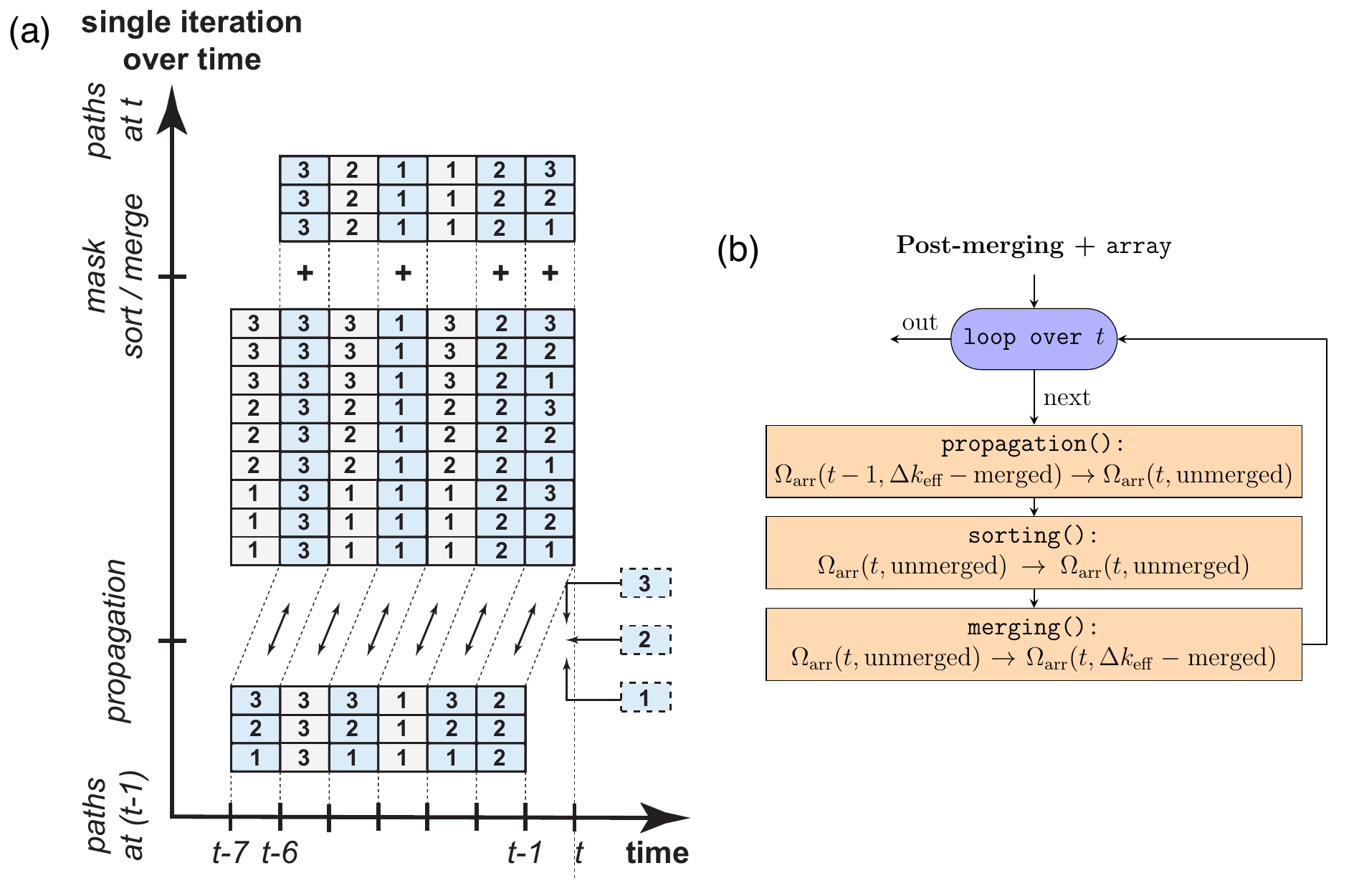}
    \caption{(a)  
    Schematic depiction of the MACGIC-QUAPI algorithm~\cite{a:Fingerhut_17} for the forward propagation of a quantum system of 3 states. Shown are three exemplary paths, as a particular evolution of the system coordinate through the memory time $\tau = \Delta k_{max}\Delta t$ with $\Delta k_{\rm max} = 6$, the considered mask of size $\Delta k_{\rm eff} = 4$ is highlighted in blue.
    Propagation from $t-1$ to $t$ involves an intermediate propagator tensor $\Omega(t,\emph{unmerged})$ of increased size
    and the  ''${\pmb{+}}$'' operation symbolizes the mask merging step of paths being identical on the mask, leading to $\Omega(t,\Delta k_{\rm eff} - \emph{merged})$.
    (b) Flowchart of subroutines   \texttt{propagation()}, \texttt{sorting()} and \texttt{merging()}.  The  \texttt{array} data structure of the programming language C was used in Ref.~\citenum{a:Fingerhut_17} to represent the $\Omega$-tensor.
    Consecutive \texttt{sorting()} and \texttt{merging()} allowed to reduce the scaling to $O(\Delta k_{\rm eff} \cdot n)$ $O(n)$ which is relevant for large $\Omega$-tensors
    (for details see text).
    }
\label{sch:macgic}
\end{scheme*}
The flowchart of Scheme~\ref{sch:macgic}b illustrates the subroutines   \texttt{propagation()}, \texttt{sorting()} and \texttt{merging()} consecutively invoked in every time step for the propagation of the reduced density matrix and the subsequent mask assisted merging of quantum paths.
Every quantum path, as a particular forward or  backward propagation evolution
of the system coordinate through memory time $\tau = \Delta k_{max}\Delta t$, is associated with a particular complex amplitude of forward and backward propagation along the path, defining its weight. 
Amplitude and path are subsumed as configurations that constitute the propagator tensor $\Omega$. 



Propagation from time step $t-1$ to $t$ involves going through all paths and propagating them one time step further which produce $\times M^2$ additional paths (corresponding to forward and backward propagation) at the current time $t$,
i.e., transiently increasing size of the propagator tensor to $\Omega(t,{\rm \emph{unmerged}})$ (Scheme~\ref{sch:macgic}a).
 The merging of paths that coincide on the coarse grained temporal grid specified by mask function of size $\Delta k_{\rm eff}$ subsequently reduces the size of the tensor. Here, among the identical path selected by the mask, 
 the paths with highest weight are retained and the complex weight of merged paths is accumulated.
Because all possible paths are considered during merging and the weights of redundant paths 
are summed up,  the trace of the reduced density matrix is preserved in the MACGIC-QUAPI method.

The merging procedure  reduces  the  number  of distinguishable paths formally to $M^{2k_{\mathrm{eff}}} \leq M^{2\Delta k_{\mathrm{max}}}$. 
The required searching within an \texttt{array} data structure is proportional to the size of the array, $O(n)$, and a simple algorithm requires $O(n^2)$ operations to perform the merging on the mask.
The consecutive use of \texttt{sorting()} via the Radix sort algorithm (requiring $O(\Delta k_{\rm eff} \cdot n)$ operations) and \texttt{merging()} routines allows to reduce the computational scaling 
to  $O(\Delta k_{\rm eff} \cdot n)$ $O(n)$.


The memory bottleneck of the MACGIC-QUAPI algorithm  is the intermediate tensor 
$\Omega(t, {\rm unmerged})$
that  formally contains $M^{2 \Delta k_{\rm eff}+2}$ 
distinguishable paths. 
A reduction of the number of paths is possible through the filtering of the  propagator functional~
\cite{Sim:JCP:2001} where 
paths with weights below a given threshold $\theta$ are not accounted for in the simulation but the efficiency of path filtering is highly problem dependent.\cite{a:Fingerhut_17,a:Fingerhut_19}

\section{Scalable Distributed Memory Implementation~} 
\label{sec:impl}

In the following we present an implementation of the MACGIC-QUAPI method that mitigates the memory bottleneck of the algorithm via the reordering of merging and propagation routines.
The novel C++ implementation numerically represents paths and their associated complex amplitude as  configuration objects \texttt{conf}.
MPI parallelization allows to exploit the distributed memory resources of multiple compute nodes.
%


\subsection{Pre-merging MACGIC-QUAPI Algorithm~} 
\label{sec:premerge}

A flowchart of subroutines of the revised algorithm is presented in Scheme~\ref{lst:merging}.
%
%
\begin{scheme*}[b!]
\includegraphics[width=0.92\textwidth]{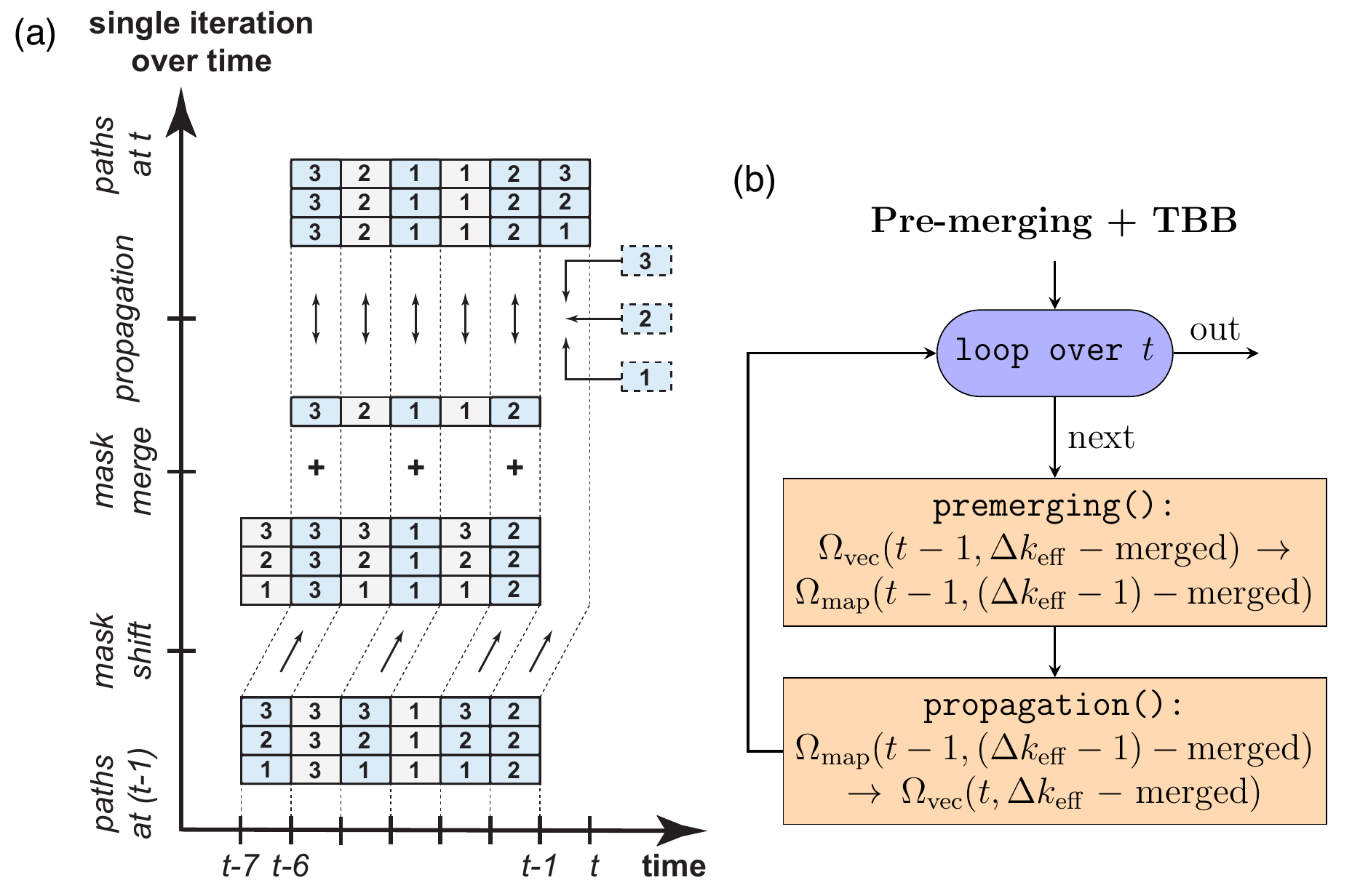}
    \caption{
     Schematic depiction of the pre-merging MACGIC-QUAPI algorithm for the forward propagation of a quantum system of 3 states. Shown are three exemplary paths, as a particular evolution of the system coordinate through the memory time $\tau = \Delta k_{max}\Delta t$ with $\Delta k_{\rm max} = 6$, the considered mask of size $\Delta k_{\rm eff} = 4$ is highlighted in blue.
    Propagation is initiated by merging step where ''${\pmb{+}}$'' symbolizes the merging operation of paths being identical on the mask, followed by propagation from $t-1$ to $t$. 
    Propagation involves an intermediate propagator tensor $\Omega(t-1,(\Delta k_{\rm eff}-1) - \emph{merged})$ of reduced size $(\Delta k_{\rm eff}-1) - \emph{merged})$ that increases to .
    $\Omega(t,\Delta k_{\rm eff} - \emph{merged})$ at time $t$.
    (b) Flowchart of subroutines \texttt{premerging()} and  \texttt{propagation()} with the corresponding input and output quantities.  
    Propagator tensors $\Omega$ are represented by the C++ data structures \texttt{tbb::concurrent\_vector} and \texttt{tbb::concurrent\_hash\_map}, respectively, implemented via  Intel Threading Building Blocks (TBB) library. The use of \texttt{tbb::concurrent\_hash\_map} data structure facilitates 
    $O(1)$ constant access time to individual configurations (for details see text).
    %
    %
}
\label{lst:merging}
\end{scheme*}
 In the new path pre-merging strategy that precedes the propagation step, 
the memory bottleneck of the original MACGIC-QUAPI algorithm is evaded. 
This is realized by initially reducing
the set of paths  $\Omega(t-1, \Delta k_{\rm eff}{\rm -\emph{merged}})$ 
via merging with a mask of reduced size $\Delta k_{\rm eff}-1$, 
generating the pre-merged 
$\Omega(t-1, (\Delta k_{\rm eff}-1){\rm -\emph{merged}})$-tensor at time step $t-1$
 that is propagated to $t$.
Upon propagation, 
the size of $\Omega(t, \Delta k_{\rm eff}{\rm -\emph{merged}})$ is recovered.
%
 %
Thus, the largest data structures in the improved algorithm are 
$\Omega(t-1, (\Delta k_{\rm eff}-1){\rm -merged})$ 
and 
$\Omega(t, \Delta k_{\rm eff}{\rm -merged})$, of size $O(M^{2(\Delta k_{\rm eff}-1)})$ and $O(M^{2 \Delta k_{\rm eff}})$, respectively, where the size of the latter can be further reduced by filtering.
The number of paths kept in memory is thus $M^2$-times reduced compared to the original 
algorithm, reducing its memory bottleneck.

 A reduced mask of size $\Delta k_{\rm eff}-1$,
 \footnote{
  The mask of reduced size $\Delta k_{\rm eff}-1$ is obtained by shifting the original mask by one time step to latter times (c.f. Scheme~\ref{sch:macgic}), i.e. ${\rm mask}[i] \to {\rm mask}[i]+1, \forall i$.  
 } 
 is used in the pre-merging operation that effectively reduces the memory time $\tau$ by one, i.e., the last time step. For '\emph{sluggish}' environments with long-time bath memory covering $\approx$ tens of time steps, the total time lag covered by the memory time is substantial and the effect of influence coefficients (eq.~\ref{eq:inf_fun}) with large time lags $d = i - j \rightarrow \Delta k_{\rm max}$  should becomes negligible for converged long-time bath memory.
 After the propagation step at time $t$,  the  tensor $\Omega(t, \Delta k_{\rm eff}{\rm -merged})$ 
 should contain the same configurations as the  original post-merging implementation. As  this is not rigorously guaranteed, numerical accuracy and stability of the pre-merging algorithm is demonstrated in Sec.~\ref{Sec:Premerge}.



\subsubsection{Data Structure and Look-up Algorithm~} 

The pre-merging MACGIC-QUAPI algorithm is implemented in C++ and  uses the concurrent version of C++ Standard Template Library (STL) serial containers as implemented in the Threading Building Blocks (TBB) library~\cite{wp:intelTBB}.
Paths and their associated complex amplitude are numerically represented as configuration \texttt{conf} objects. The abstract set of configurations 
is denoted as $\Omega$-tensor.
Propagator tensors $\Omega$ in subroutines \texttt{premerging()} and \texttt{propagation()} are represented two different data structures, the C++ data structure \texttt{tbb::concurrent\_vector} ($\Omega_{vec}$) and an intermediate data structure \texttt{tbb::concurrent\_hash\_map} ($\Omega_{map}$), respectively (Scheme~\ref{lst:merging}b, a detailled discussion of data structures in given in the \ToDo{Supporting Information (SI)}).
    
The pre-merging algorithm exploits the built-in 
search, insertion, and deletion properties of the associative map container capitalizing on the advantage that operations can be performed concurrently by different TBB threads. In particular hash maps $\Omega_{\rm map}$ (consisting of \texttt{(key; value)} pairs) provide the crucial property of a constant access time to any \texttt{value} by its \texttt{key}, i.e., searching, inserting, and erasing scale as $O(1)$, regardless of the size of the  $\Omega_{\rm map}$-tensor. 
Thus, the  $O(\Delta k_{\rm eff} \cdot n)$ scaling of the sorting step in the original algorithm can be circumvented (cf. Scheme~\ref{sch:macgic}).

The records \texttt{value} of an associative map are distributed in heap memory in no specific order, so the hash map grows dynamically in an efficient way with an increase in the number of paths.
%
%
%
 A particular \texttt{value = (conf; sum)} is defined by 
a particular configuration \texttt{conf = (paths; weight)} and the running sum \texttt{sum} of the weight if paths are merged. 
%
In Sec.~\ref{Sec:Premerge} we numerically demonstrate that competitive performance  for such data structure  can be achieved compared to  simpler \texttt{array} data structure, even for the significantly slower traversal times of associative maps.
\begin{scheme*}[ht!]
\includegraphics[width=0.98\textwidth]{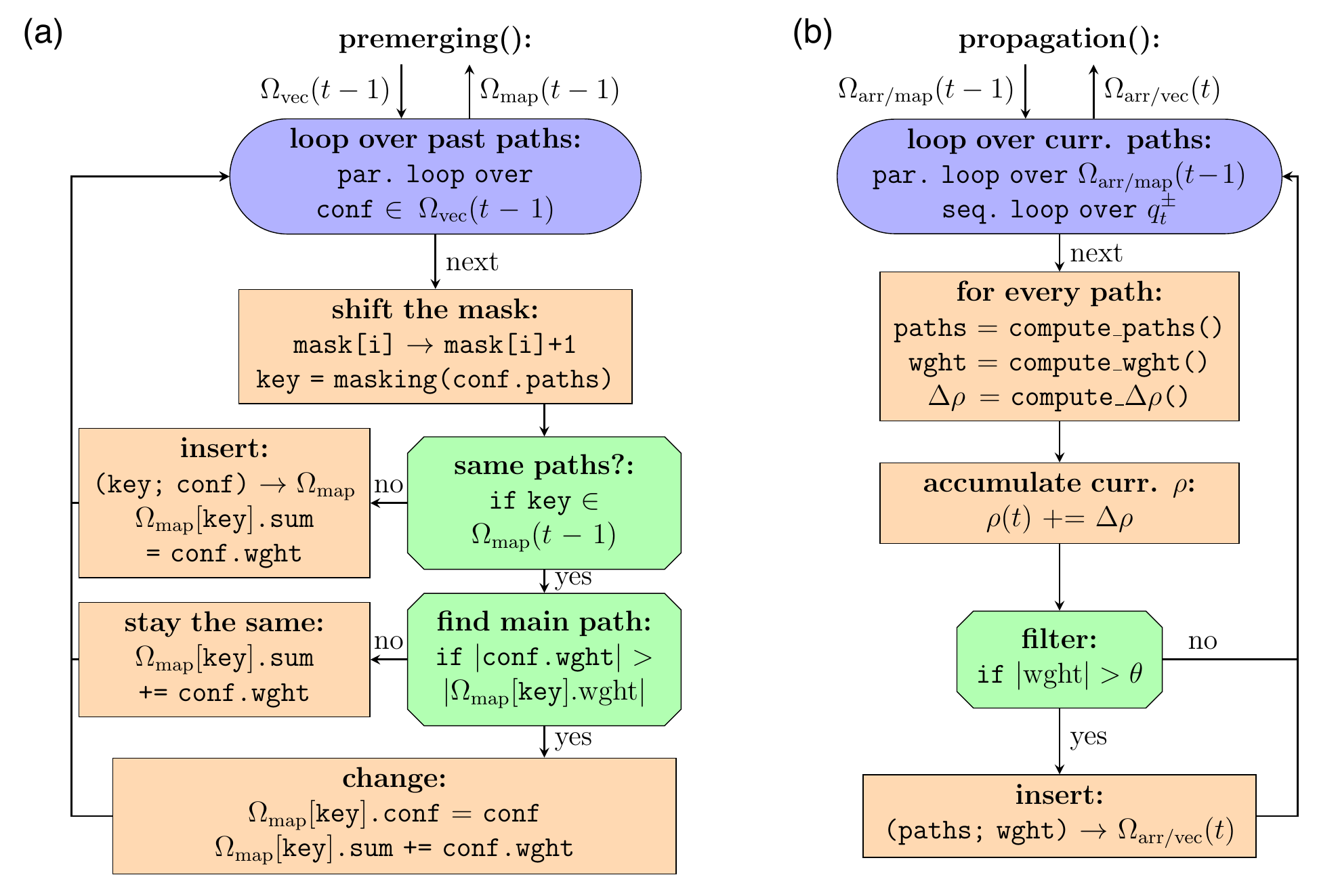}
    \caption{
    Flowcharts for the \texttt{premerging()} (a) and \texttt{propagation()} routines (b) introduced in Scheme~\ref{lst:merging}. Both subroutines 
    use the data structures \ToDo{$\Omega_{\rm vec}$ and $\Omega_{\rm map}$}
    to represent the propagator tensor $\Omega$,
    i.e., data structures \texttt{tbb::concurrent\_vector} and \texttt{tbb::concurrent\_hash\_map}, respectively.
    %
    \texttt{conf} is an C++ object representing a single configuration made up of the pair \texttt{(paths; wght)}. The $\Omega_{\rm map}$ structure stores unordered pairs \texttt{(key; (conf; sum))}, where  \texttt{key} uniquely points to the \texttt{(conf; sum)} pair, and \texttt{sum} keeps the running sums of weights for the merged paths. 
    Host data structures and their attributes are separated by a dot, i.e., \texttt{$\Omega_{\rm map}[$key$]$.wght} reads as weight corresponding to the \texttt{key}, residing in the $\Omega_{\rm map}$ data structure. Attributes without host are related to the running index of loops.
    }
\label{lst:propagation}
\end{scheme*}

\subsubsection{Pre-merging~} 
\label{sec:premergeRoutine}

In the \texttt{premerging()} routine (Scheme~\ref{lst:propagation} (a)), we iterate through configurations of the $\Omega_{\rm vec}(t-1)$-tensor at time step $t-1$.
For every configuration in $\Omega_{\rm vec}(t-1)$ the mask of size $(\Delta k_{\rm eff}-1)$ is applied and we retrieve the \texttt{key} of the masked configuration. We further check if the unique key already exists in the new $\Omega_{\rm map}(t-1)$-tensor. 
 If such \texttt{key} is absent, we insert the \texttt{key} and the corresponding \texttt{conf} to the hash map. 
 If the \texttt{key} is present, i.e., an identical masked but otherwise distinguishable path exists, we perform a comparison if the weight \texttt{wght} (measured as absolute value) of the new path is larger 
 than that of the old path contained in the hash map.  
 If this is the case, the old path is replaced with the new one by 
 updating the configuration corresponding to the \texttt{key} in the hash map. 
 Remember, that only the path with the largest weight is saved and employed for further propagation. 
 
 On the other hand, if the weight of the new path is smaller that the old path already contained in $\Omega_{\rm map}(t-1)$, the weights of identical masked paths 
 are added up.
  That is why together with the weight of every configuration, we keep track of the summed up value \texttt{sum}. 
  %
 %
 %
 Note that, since the search and insertion of a pair into the hash map take constant time $O(1)$, the overall scaling of the premerging routine is linear $O(n)$. 


\subsection{Distributed Memory Parallelization~} 
\label{sec:dist_mem}

In this section, we present the distributed memory parallelization of the MACGIC-QUAPI method.
 The 
implementation makes accessible the compute resources of multiple compute nodes which is particularly appealing for the substantial  memory requirements of the QUAPI method. 
An efficient algorithm has to  perform the work loads of path merging and propagation as independent as possible on an individual compute node.
This requirement is realized by leveraging the threads-based parallelization of the pre-merging algorithm to a multi-node parallel execution via message passing interface (MPI) 
with the consecutive intra- and inter-node pre-merging operations.


\subsubsection{General Code Structure~} 

Scheme~\ref{sch:mpi_scheme} shows the  global code structure of the distributed memory parallel 
MAGCIC-QAUPI method. A guiding principle of the implementation is provide the maximum amount of memory of each compute node
by equally distributing the paths of the $\Omega$-tensor.
Each node then performs the path pre-merging and propagation steps independently on a set of assigned paths. 
\begin{scheme*}[t!]
\includegraphics[width=\textwidth]{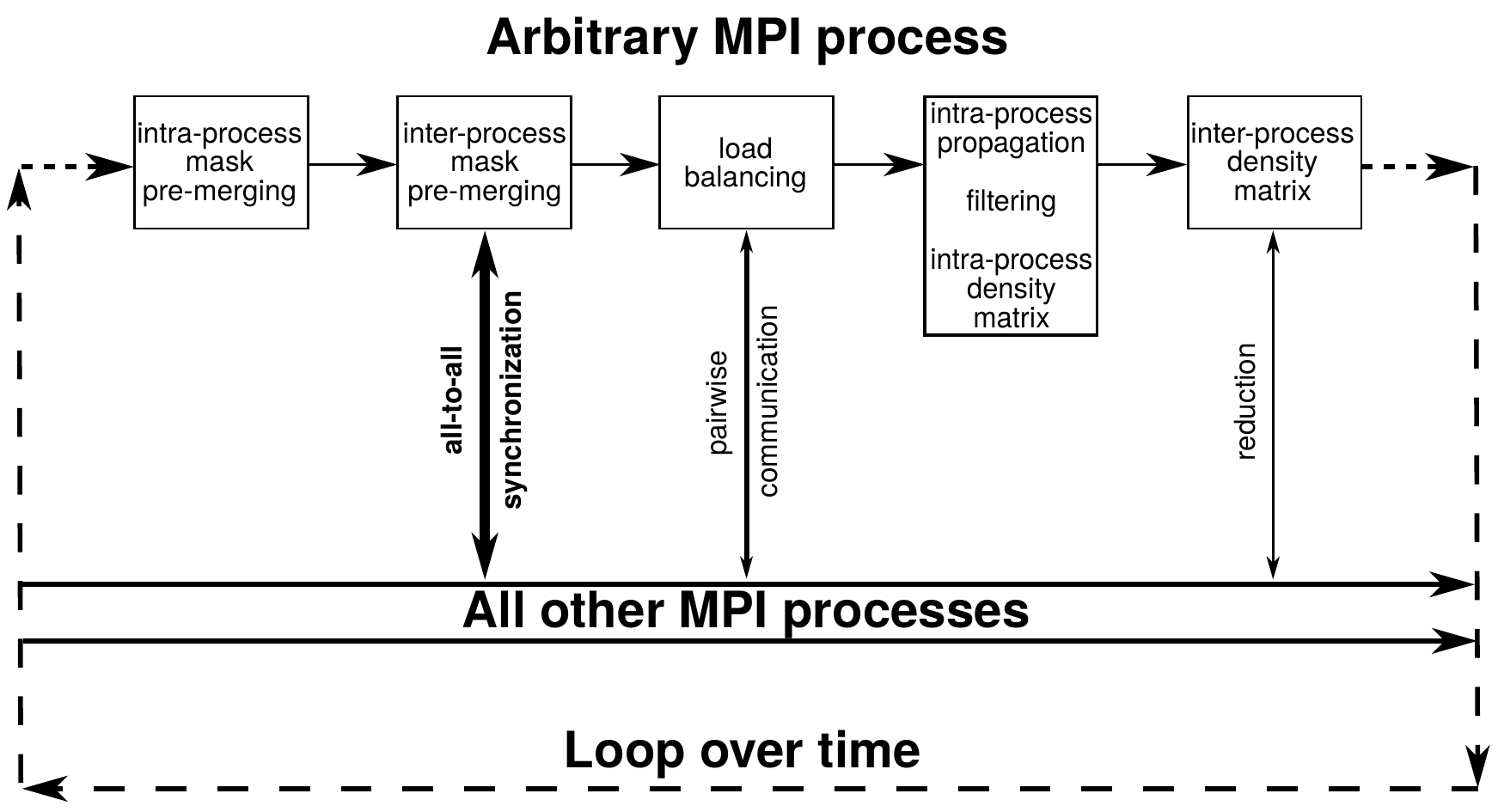}
\caption{The communication infrastructure between MPI processes during one propagation step. 
}
\label{sch:mpi_scheme}
\end{scheme*}

Initially at every propagation step, 
the pre-merging of the $\Omega_{\rm vec}(t-1)$-tensor is performed, using the  mask of size $(\Delta k_{\rm eff}-1)$. For parallel execution, the  pre-merging operation is split into two parts: 
it is first performed independently within each MPI process using the pre-merging  routine described in Sec.~\ref{sec:premerge} (\textit{intra-process pre-merging}). The routine operates on all paths that reside within memory of a particular process.
However, for different processes with separate memory, there  may still exist identical paths that are consolidated by 
an  \textit{inter-process pre-merging} routine.
 \emph{Per se},  identical paths may reside in all processes, requiring an all-to-all communication model for \textit{inter-process pre-merging}. 
 Thus, the effective implementation of this 
 routine is crucial for the overall performance (c.f. Sec.~\ref{sec:InterNode}).
Moreover, merging 
brings the processes out of balance. We thus implemented \textit{load balancing} between the individual processes 
in a two-step process with a pairwise communication model (c.f. Sec.~\ref{sec:loadbalance}).

Subsequently, 
propagation of the density matrix is performed via \textit{intra-process propagation}, \textit{filtering}, and \textit{intra-process density matrix} with the \textit{propagation} routine of Scheme~\ref{lst:propagation}, each operating independently on the
pre-merged $\Omega_{\rm map}(t-1)$ tensor.
The generated $\Omega_{\rm vec}(t)$-tensor at time $t$ is of size $\Delta k_{\rm eff}$. 
%
Finally, 
the density matrices at time $t$, residing in different processes, are added up (\textit{reduction}) to obtain the final density matrix.
 Since the size of the density matrix is $M \times M$, the inter-process communication of this generally rather small matrix is a high-speed operation.

\subsubsection{MPI Implementation of Inter-process Path Merging~} 
\label{sec:InterNode}
The parallel distributed memory algorithm outlined in Scheme~\ref{sch:mpi_scheme} requires to synchronize the configurations in every process which is an all-to-all communication pattern. Efficient inter-process path merging thus poses a particular challenge of the algorithm.
%
%
We thus implemented a strategy (Scheme~\ref{lst:inter_merg_intel}) that balances concurrency and reduces the memory overhead. 
The algorithm iterates sequentially through the processes and  the current process (called sender) broadcasts a local set of configurations $\Omega_{\rm loc}$ residing in the current process to all other processes (called receivers).
The receivers obtain a set of configurations 
from the sender and perform the main workload of the inter-process path merging routine in a TBB-parallelized loop. 
Receivers are classified into two groups:
the ones that have already taken the role of the sender ($processes < sender$) and others that have not yet acted as sender ($processes > sender$).
A set of configurations received from the sender (
(denoted $\Omega_{\rm rem}$ on receivers) is then compared to the local set of configurations.
The latter can either be the original, unchanged set of configurations $\Omega_{\rm loc}$ or  an already changed set of local configurations $\Omega_{\rm loc}^*$, 
 depending if the receiver process has already acted as sender ($processes > sender$ or $processes < sender$, respectively).
In this way, the locally stored set of configurations are synchronized with  
the received set of configurations for every process with minimal memory overhead.
The overall memory requirement is $\sim 2/N_{\rm proc} \times T$ per process, where $N_{\rm proc}$ is the number of spawned processes. 
%
%
%
%
\begin{scheme*}
\begin{tikzpicture}[node distance=1.5cm]
\node (seq_loop) [loop] {Sequential loop: sender = $0 ... N_{\rm proc}$};
\node (seq_node) [io, below of=seq_loop, yshift=0.5cm, text width=0cm, text height=0cm] {};
\node (start_mid) [io, below of=seq_node, yshift=0.5cm] {\textbf{process $\mathbf{=}$ sender}};
\node (start_left) [io, left of=start_mid, xshift=-4cm] {\textbf{processes $\mathbf{<}$ sender}};
\node (start_right) [io, right of=start_mid, xshift=4cm] {\textbf{processes $\mathbf{>}$ sender}};
\node (send_mid) [comm, below of=start_mid, text width=4.8cm] {Broadcast local chunk of configurations $\Omega_{\rm loc}$};
\node (get_left) [comm, text width=4.8cm] at (start_left |- send_mid) {Get remote chunk of configurations $\Omega_{\rm rem}$};
\node (get_right) [comm, text width=4.8cm] at (start_right |- send_mid) {Get remote chunk of configurations $\Omega_{\rm rem}$};
\node (proc_mid) [tbb_process, below of=send_mid, text width=4.8cm, yshift=-2cm] {\textbf{Merge $\mathbf{\Omega_{\rm loc}}$ and $\mathbf{\Omega_{\rm aux}}$:} \begin{itemize}[leftmargin=*] \item Replace original local weights in $\Omega_{\rm loc}$ with values accumulated in $\Omega_{\rm aux}$: $\Omega_{\rm loc} \to \Omega_{\rm loc}^*$ \item Delete $\Omega_{\rm aux}$ \end{itemize}};
\node (proc_left) [tbb_process, below of=get_left, text width=4.8cm, yshift=-2cm] {\textbf{Merge $\mathbf{\Omega_{\rm loc}^*}$ and $\mathbf{\Omega_{\rm rem}}$:} \begin{itemize}[leftmargin=*] \item Compare weights of identical paths in $\Omega_{\rm loc}^*$ and $\Omega_{\rm rem}$ \item Keep in $\Omega_{\rm loc}^*$ only paths with largest weights \end{itemize}};
\node (proc_right) [tbb_process, below of=get_right, text width=4.8cm, yshift=-2cm] {\textbf{Accumulate $\mathbf{\Omega_{\rm aux}}$:} \begin{itemize}[leftmargin=*] \item Compare weights of identical paths in $\Omega_{\rm loc}$ and $\Omega_{\rm rem}$ \item Accumulate weights in $\Omega_{\rm aux}$ \end{itemize}};
\begin{scope}[on background layer]
\node (MPI_mid) [draw,dashed,black,rounded corners,fill=green!50,fit=(start_mid)(proc_mid)]{};
\node (MPI_left) [draw,dashed,black,rounded corners,fill=green!50,fit=(start_left)(proc_left)]{};
\node (MPI_right) [draw,dashed,black,rounded corners,fill=green!50,fit=(start_right)(proc_right)]{};
\node (end_node) [io, below of=MPI_mid, yshift=-2.8cm, text width=0cm, text height=0cm] {};
\node (end_end_node) [io, below of=end_node, yshift=1.0cm, text width=0cm, text height=0cm] {};
\end{scope}
\draw [carrow] (send_mid.west) -- (get_left.east);
\draw [carrow] (send_mid.east) -- (get_right.west);
\draw [arrow] (seq_loop.south) -- (MPI_mid.north);
\draw [arrow] (seq_node.west) -| (MPI_left.north);
\draw [arrow] (seq_node.east) -| (MPI_right.north);
\draw [arrow] (MPI_left.south) |- (end_node.west);
\draw [arrow] (MPI_right.south) |- (end_node.east);
\draw [arrow] (MPI_mid.south) -- (end_end_node.north);
\draw [arrow] (send_mid.south) -- (proc_mid.north);
\draw [arrow] (get_left.south) -- (proc_left.north);
\draw [arrow] (get_right.south) -- (proc_right.north);
\end{tikzpicture}
\caption{
Single iteration flowchart of the inter-process mask merging routine. Red blocks and horizontal lines mark the MPI communication routines and data flows, respectively. Yellow rectangles encompass computationally extensive TBB-parallelized for-loops, while the green rectangles denote MPI-parallelized regions. $N_{\rm proc}$ is the total number of the MPI processes and $\Omega_{\rm loc}$ denotes an original (unchanged) set of paths and associated weights assigned to  each process. 
The sender broadcasts  the $\Omega_{\rm loc}$-tensor that is received as $\Omega_{\rm rem}$. 
Modification of the original weights for paths identical on the mask results on the  $\Omega_{\rm loc}^*$ data structure, the accumulating tensor $\Omega_{\rm aux}$ is introduced as compact auxiliary representatoin (see text for details).
}
\label{lst:inter_merg_intel}
\end{scheme*}

The main purpose of the group of receivers with $processes > sender$ 
is to accumulate the weights of identical paths 
that are present in the local $\Omega_{\rm loc}$-tensor. 
As the original $\Omega_{\rm loc}$-tensor is still required for transmission via the MPI library  when the process becomes sender ($processes = sender$), identical paths in $\Omega_{\rm loc}$ cannot just be replaced. 
We thus use an auxiliary data structure $\Omega_{\rm aux}$ that is merged with the $\Omega_{\rm loc}$  data structure after the broadcasting of the original local set of configurations. Similarly, the original weights are replaced with the accumulated sums, generating the locally merged $\Omega_{\rm loc}^*$-tensor.
For optimal performance, the local tensors $\Omega_{\rm aux}$ and $\Omega_{\rm loc}^*$ are represented as hash maps, while the transmitted tensors $\Omega_{\rm rem}$ are implemented as \texttt{array}. 
While Scheme~\ref{lst:inter_merg_intel} depicts $\Omega_{\rm loc}$ as single data structure for simplicity, the 
actual implementation the $\Omega_{\rm loc}$ consists of two objects: one hash map and one \texttt{array} (see  \ToDo{SI} for numerical implementation details of the inter-process path merging routine)

Memory requirements are minimized by using a compact representation that replaces each path with the masked path \texttt{key}. 
instead of storing the full tensors $\Omega_{\rm aux}$- and $\Omega_{\rm loc}$.
This allows to continuously update the locally  merged $\Omega_{\rm loc}^*$-tensor via the comparison with the received set of of configurations $\Omega_{\rm rem}$
for $processes < sender$ receivers

\subsubsection{Load Balancing~} 
\label{sec:loadbalance}

The asynchronous operation of inter-process path merging brings the workload of individual processes out of balance. 
\textit{Load balancing} is thus implemented in a two-step process:
First, the total number of configurations is calculated and divided by the number of processes $N_{\rm proc}$ to obtain the average number of configurations per process.
 Two lists are then created 
 containing processes with less or more configurations than the average where the imbalances have arisen from the \textit{inter-process process pre-merging}. 
 Deficient and overcrowded processes are subsequently combined pairwise to obtain a route map, i.e., a map containing sending processes, receiving processes, and the number of configurations to transmit.
 
  In the second part, the route map is then used to 
  choose a given number of 
  configurations from the sender, pack them into the MPI message, and send them to the receiver. Such a model corresponds to a  pairwise communication pattern.
Moreover, the number of transmitted configurations is given by the difference of the number of actual  configurations residing in a  particular process and the average. Such deviations are much smaller than the total number of configurations residing in any particular process, the \textit{load balancing} routine is thus much faster and substantially less communication intense than \textit{inter-process merging}.

\section{Results and Discussion}

\subsection{Numerical Accuracy and Look-up Algorithm}
\label{Sec:Premerge}

We start by  demonstrating  the numerically accuracy of the mask assisted pre-merging algorithm introduced in Sec.~\ref{sec:premerge}
and pointing out advantages 
of the fast look-up of paths using the \texttt{tbb::concurrent\_hash\_map} data structure.
Numerical experiments are initially performed on a single compute node 
and compared to reference simulations described in Refs.~\citenum{a:Fingerhut_17, a:Fingerhut_19} that used the original  post-merging algorithm.
The latter implements a 
look-up of paths arranged consecutively in an \textit{array} data structures and 
 shared-memory parallelization with the openMP interface 
 (see \ToDo{SI} for a technical details). 
 
 

\begin{figure*}
\includegraphics[width=0.49\textwidth]{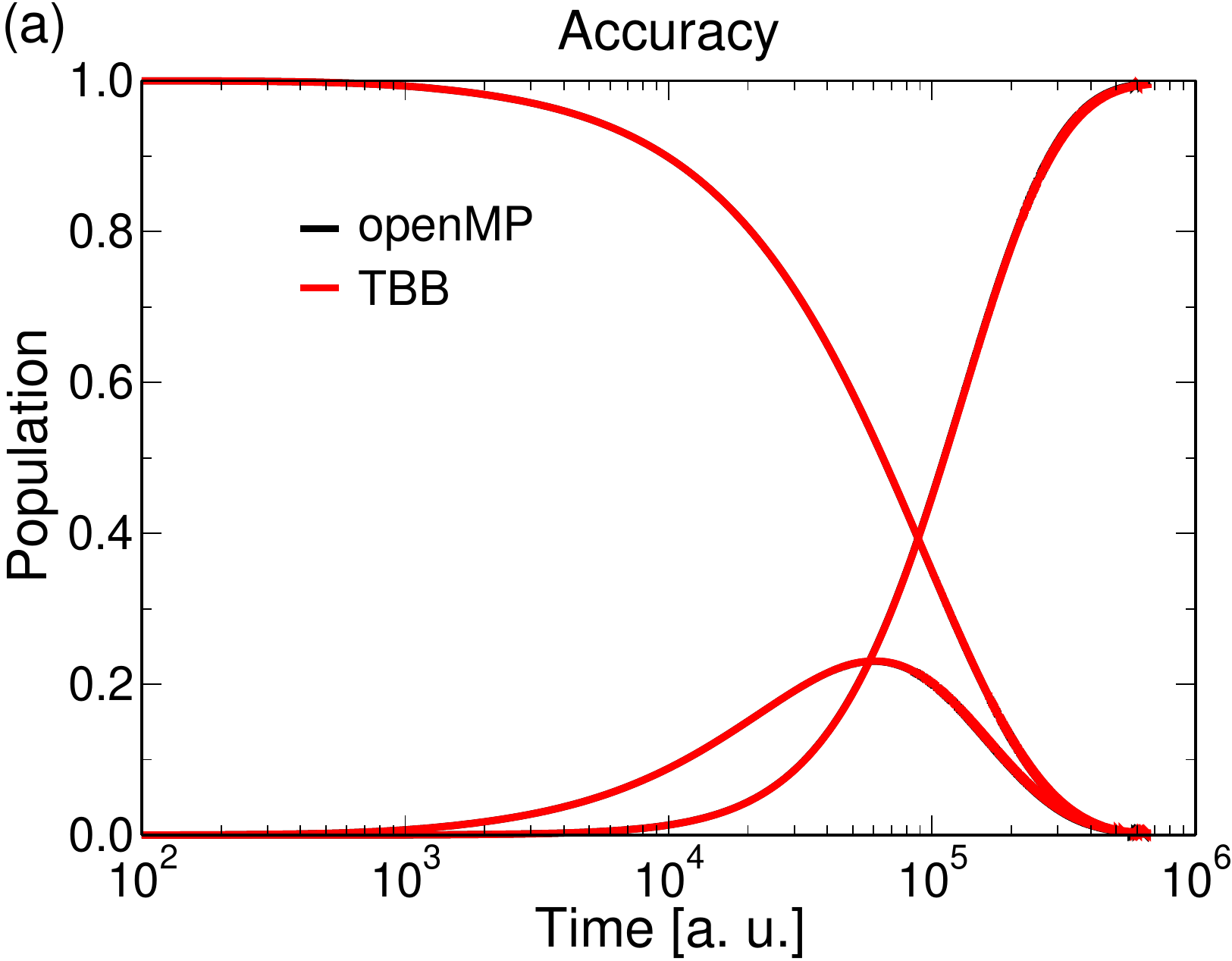}
\includegraphics[width=0.49\textwidth]{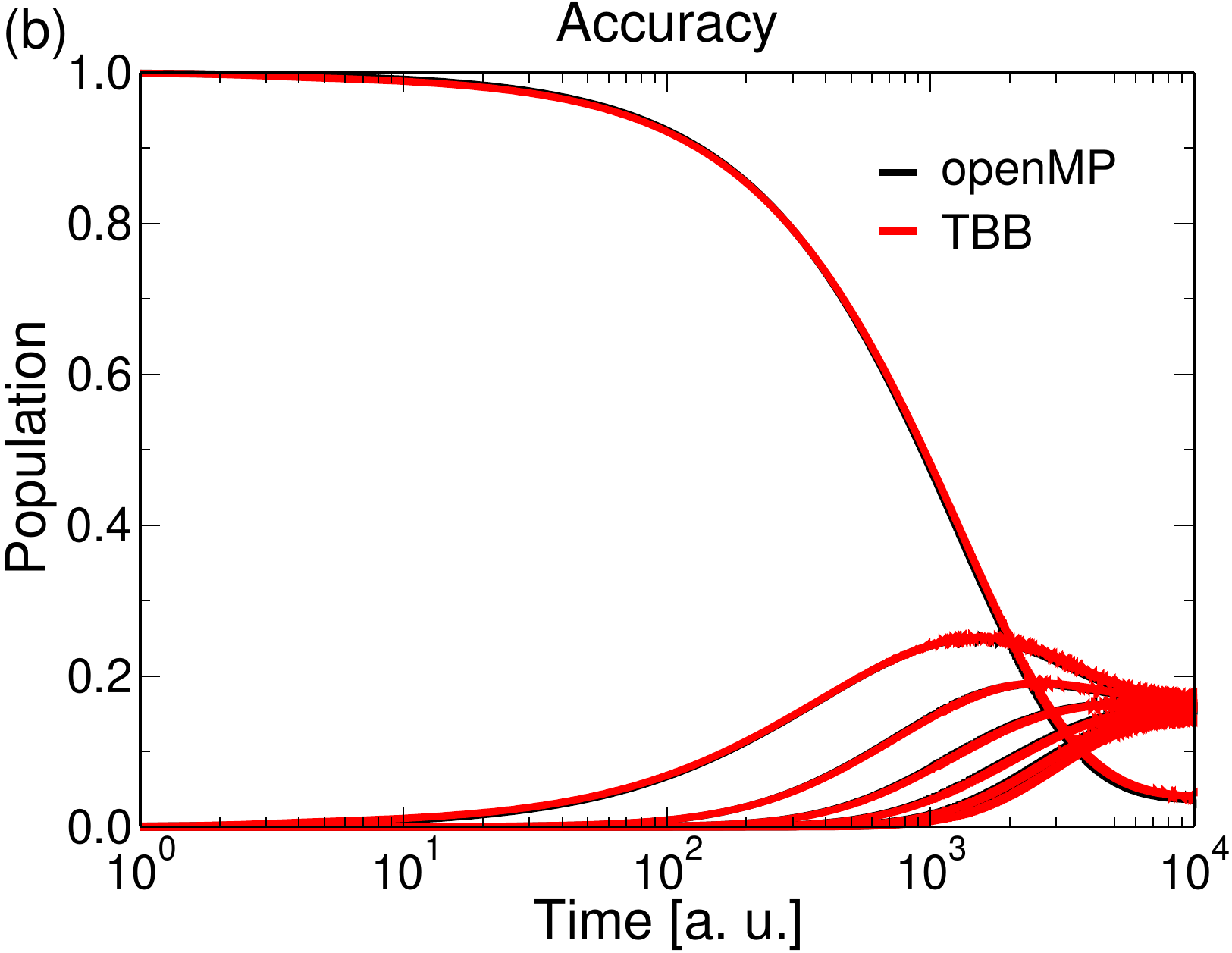}
\\
\includegraphics[width=0.49\textwidth]{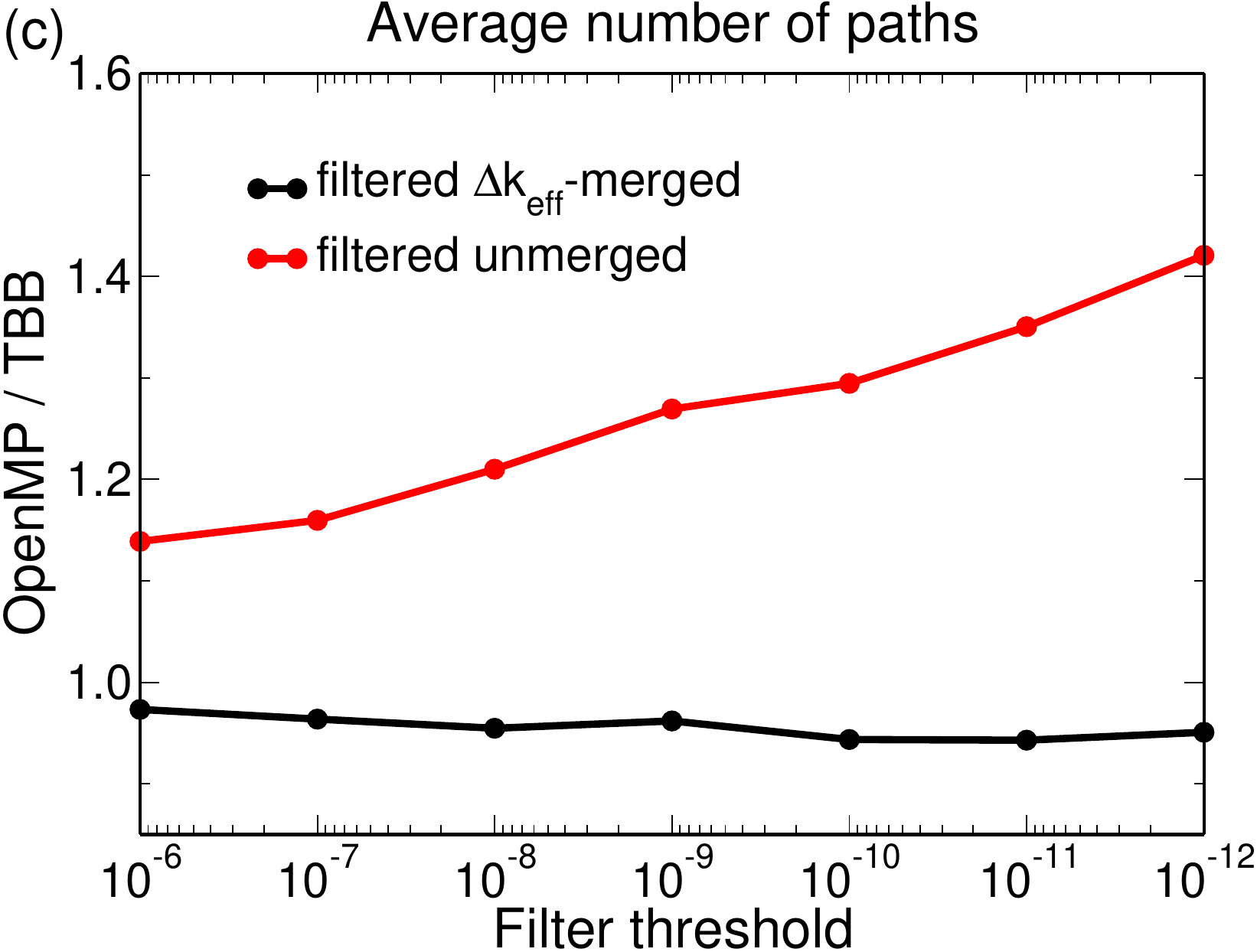}
\includegraphics[width=0.49\textwidth]{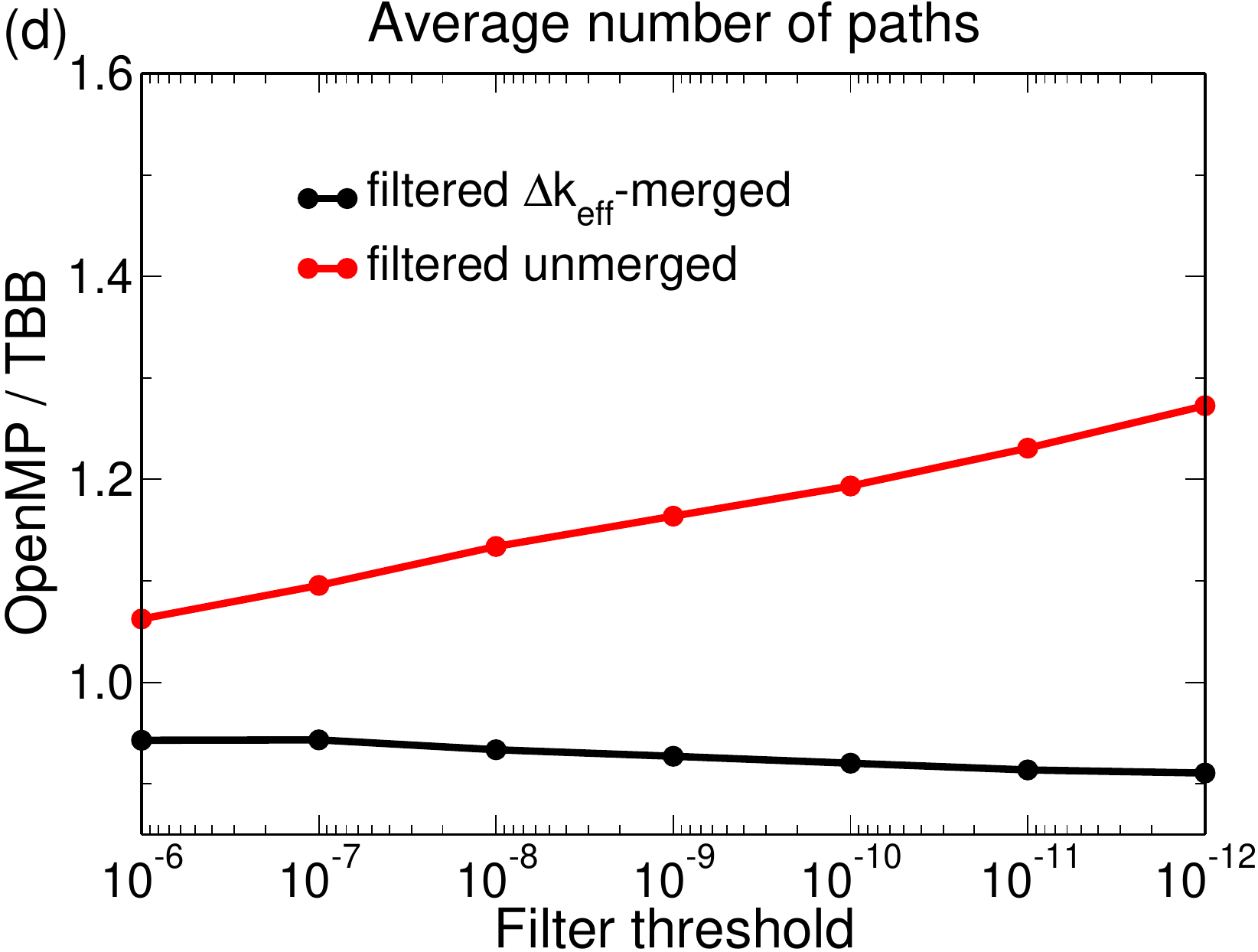}
    \caption{
    Non-equilibrium dynamics of the $3$-state system (a) and $7$-state system (b), panels (c) and (d) provide the corresponding ratios of the number of paths used on average for propagation for varying filter thresholds $\theta$.
     ``TBB'' and ``openMP'' marks the new pre-merging and original prost-merging implementations, respectively. 
     Filter threshold $\theta = 10^{-6}$ corresponds to $\sim 10^4$ paths for both reference systems,  $\theta = 10^{-12}$ corresponds to $2 \cdot 10^7$ and $4 \cdot 10^7$ paths for the $3$- and $7$-state system, respectively.
}
\label{fig:tbb_vs_openmp}
\end{figure*}

Two  benchmark models, i.e., bridge mediated electron transfer~\cite{a:Makri_97} (denoted ``$3$-state system'')  and bath assisted long range charge transfer along a chain of seven sites~\cite{a:Fingerhut_17}  (denoted ``$7$-state system'') serve to illustrate the numerical accuracy to the implementation.
Figure~\ref{fig:tbb_vs_openmp}~(a, b) demonstrates that for both models,  the numerical exact reference dynamics is reproduced with excellent accuracy with the new pre-merging algorithm (max. population
deviations  $<$ $0.001$ and $0.01$ for $3$- and $7$-state systems, respectively).
Similar high accuracy is obtained for all benchmark simulations presented in Ref.~\citenum{a:Fingerhut_17}, spanning  a wide range of physical parameter space and dynamical regimes, i.e.,  coherent excitation energy transfer in a dimeric Frenkel exciton models and the Fenna-Matthews-Olson complex (see \ToDo{SI} for dynamics). 


Memory requirements of the path pre-merging algorithm are analyzed in Fig.~\ref{fig:tbb_vs_openmp}~(c,d). 
%
%
%
For propagation,  the new pre-merging implementation selects slightly more relevant paths (ratio $~ 0.9-1$, c.f. black lines in Fig.~\ref{fig:tbb_vs_openmp}~(c,d))  with a weak dependence on the filter threshold $\theta$.
Nevertheless, the intermediate data structure of the pre-merging algorithm is significantly smaller, reducing the memory bottleneck compared to the original  ``openMP'' implementation.
Depending on the filter threshold $\theta$, path ratios range from $1.14$ ($\theta=$$10^{-6}$) to  $1.42$ ($\theta$=$10^{-12}$) demonstrating the smaller memory footprint of the pre-merging scheme.
While the memory gain in the pre-merging algorithm  is strongly filter and task specific, no crossover of  the red line and black lines in Figs.~\ref{fig:tbb_vs_openmp}~(c,d) is observerd, i.e., the ratio of path numbers approaches  one  even in unfavorable scenarios. 

The used \texttt{tbb::concurrent\_hash\_map} data structure introduces a memory overhead compared to the C \texttt{array}, as the hash map stores the 
path \texttt{key} in addition to forward and backward paths and the associated weights. In the case of a large mask ($\Delta k_{eff} \rightarrow \Delta k_{max}$) the additional  memory requirements  may reduce the memory advantage of the pre-merging algorithm but the new implementation is expected to be especially suited for long-time bath memory of 'sluggish' environments that require a reasonably small mask.

\begin{figure}
\includegraphics[width=0.49\textwidth]{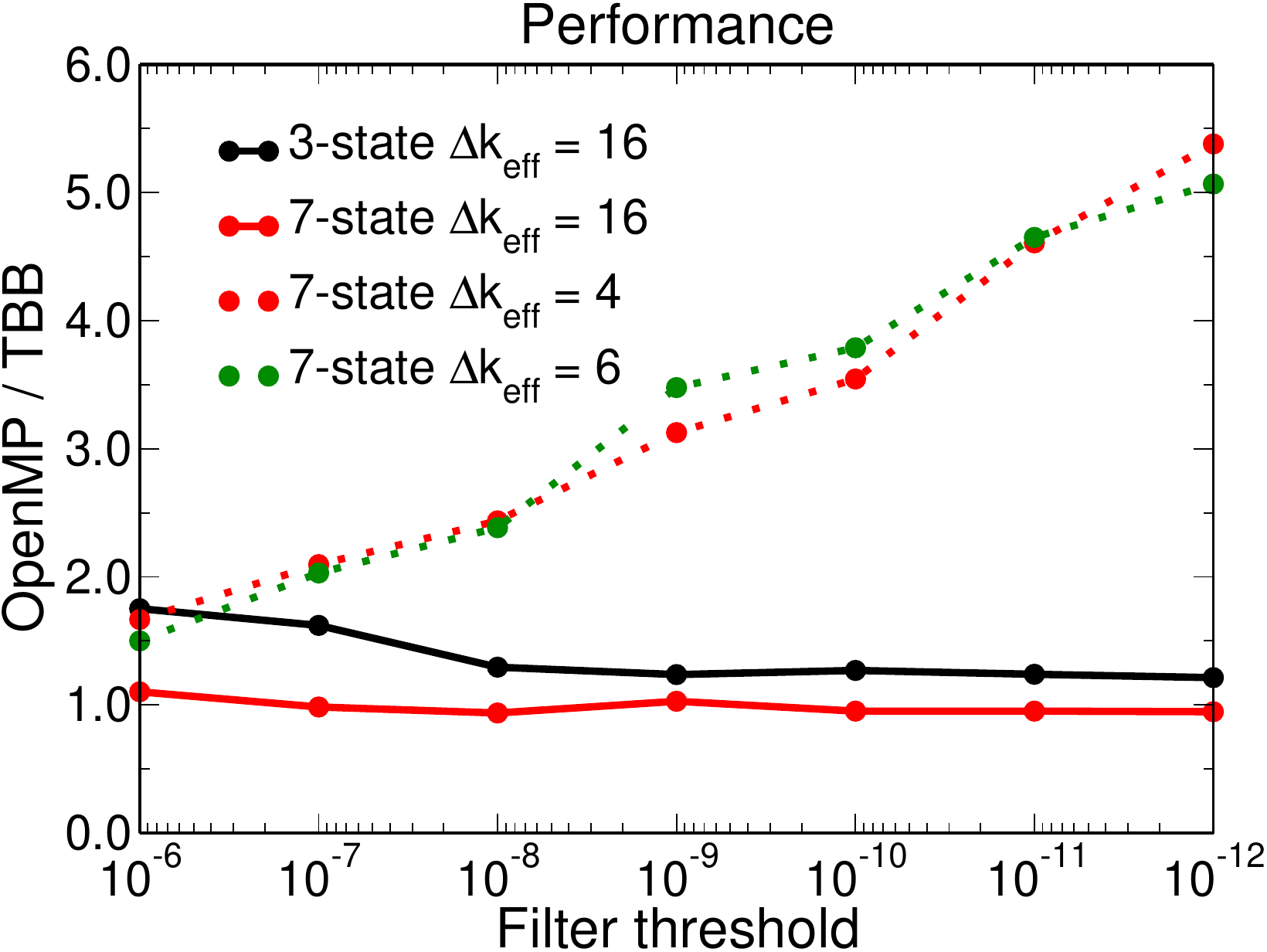}
    \caption{Relative timing with respect to filter threshold $\theta$ and variation of the mask size. In all simulations $\Delta k_{\rm max} = 32$ time steps.}
\label{fig:performance}
\end{figure}
 Figure~\ref{fig:performance}~compares relative timings of the post- and pre-merging implementation. For the $3$-state system we find that the pre-merging implementation is up to $1.75$ times faster than the original post-merging implementation (black line, $\theta$ =  $10^{-6}$). Such speedup is observed despite the slower dynamical memory allocation in heap memory, compared to the statically allocated stack memory of the original post-merging implementation.
 The reasons for the observed performance gain are twofold: First, the \texttt{tbb::concurrent\_hash\_map} data structure allows for the efficient 
 look-up of paths by avoiding  the sorting step, thereby improving the numerical performance of pre-merging algorithm.
 Second, the TBB-thread based parallelization of the pre-merging algorithm yields a better core balance via the TBB task scheduler compared to the prior openMP parallelization. This effect becomes apparent for large filter thresholds ($\theta =$$10^{-6}$), while for small filter thresholds ($\theta =$$10^{-12}$) the increase in the number of paths  decreases the core imbalance. 

Comparing the numerical performance of the pre- and post-merging algorithms for the $7$-state system, we observe a strong dependence on the employed mask size. 
For a large mask size (red solid line in Fig.~\ref{fig:performance}, $\Delta k_{\rm eff}=16$), a large number of paths is considered (ranging from $10^4$ for $\theta = 10^{-6}$  to $4 \cdot 10^7$ for $10^{-12}$) that reduces  the relative performance advantage of the  pre-merging algorithm and the relative timing ratio   approaches $\sim 1$.
For reduced mask sizes (dashed red and green lines in Fig.~\ref{fig:performance}, $\Delta k_{\rm eff}=4$ and $6$, respectively), the  pre-merging implementation is significantly faster for all investigated $\theta$.

The varying relative numerical efficiency of the pre- and post-merging implementations are rooted in the different paths look-up algorithms. 
The Radix sort algorithm used in the post-merging implementation becomes particularly efficient for partially sorted arrays between consecutive time steps, i.e., large mask sizes (Fig.~\ref{fig:performance}, $\Delta k_{\rm eff}=16$).
Sorting is avoided in the hash-table based look-up of the pre-merging algorithm, therefore the performance 
does not depend on a particular order of paths in memory and the algorithm becomes particularly efficient for smaller, sparse and more irregular masks that pose challenges to the post-merging implementation.
%
For both small masks $\Delta k_{\rm eff}=4$ and $6$  (cf. dashed red and green lines in Fig.~\ref{fig:performance}) the relative timing ratio  grows monotonically with $\theta$ as the Radix sort algorithm scales linearly with the number of paths while the ash-table based look-up provides constant access time.

\subsection{Performance Analysis of the Distributed Memory Parallelization}
\label{sec:DistrMem}
\begin{figure*}
\includegraphics[width=0.49\textwidth]{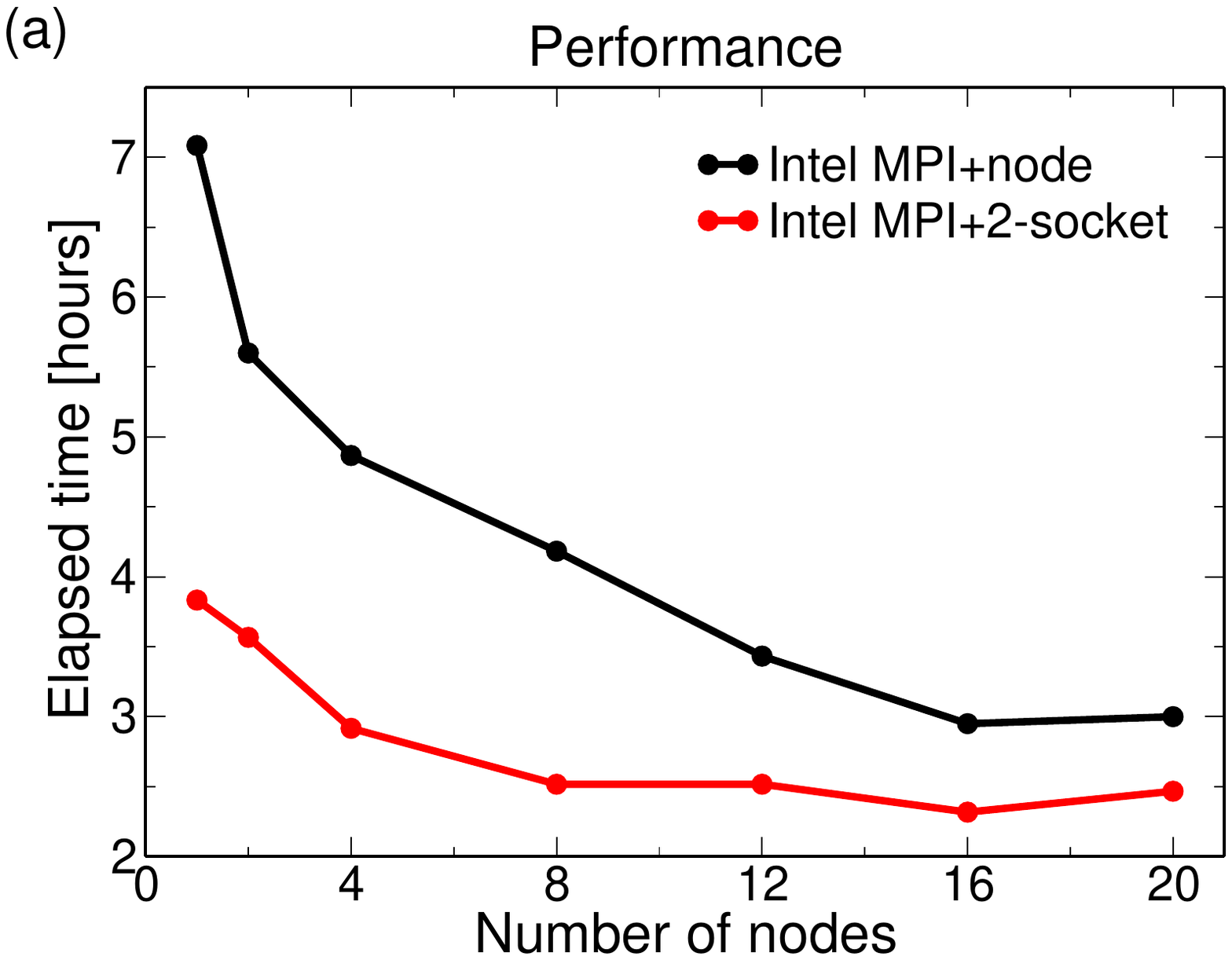}
\includegraphics[width=0.49\textwidth]{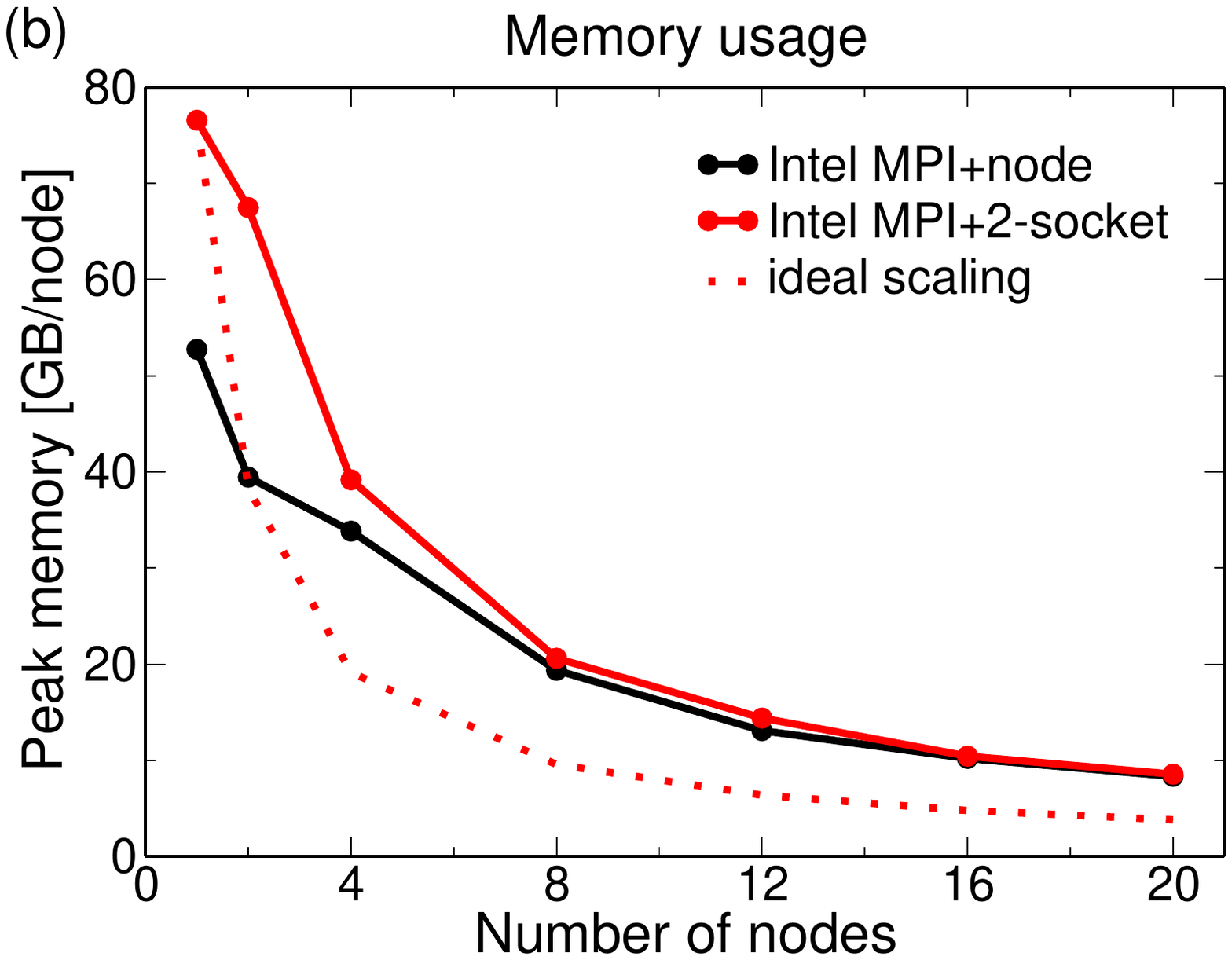}
    \caption{
    Numerical performance (a) and peak memory usage (b) for varying number of compute nodes.
    The model system  is a two level system with Hamiltonian $H_{\rm S} = \frac{\Delta}{2} \sigma_x$ 
    and diagonal coupling via $\sigma_z$ to an Ohmic bath ($J(\omega) = \gamma \pi^{-1} \, \omega \, \EXP{-\omega/\omega_c}$, where $\gamma = 1/16$ and $\omega_c = 10 \Delta$ at temperature $kT = 0.2 \Delta$ and ($\sigma_{x/z}$ being Pauli matrices). 
    Time propagation is done with a time step $dt = 0.3 \Delta^{-1}$ until $t_{\rm max} = 35 \Delta^{-1}$, the free parameter $\Delta$ was fixed at $200$~cm$^{-1}$. 
    Simulations were performed in the full QUAPI regime, i.e., no mask merging or filtering have been employed. The bath memory time $\Delta k_{max}$ is $14$ time steps, resulting in $2^{28} \sim 2.7 \cdot 10^8$ paths. 
    The ideal scaling curve (dots) was obtained by dividing the single node peak memory value by the corresponding number of nodes.}
\label{fig:intel_MPI}
\end{figure*}

The numerical behavior of the pre-merging algorithm in combination with the distributed memory parallelization using the MPI protocol is analyzed in Fig.~\ref{fig:intel_MPI}~(a,b), showing the numerical performance and memory scaling with respect to the number of compute nodes.
Two MPI process pinning schemes are considered:
 either one MPI process per node is used (i.e., one MPI process per two sockets, denoted ``node'') or one MPI process per socket  (i.e.,  two MPI processes per node, denoted ``$2$-socket'') is used. For both pinning schemes, the task based shared-memory TBB parallelization is exploited over the processor cores. 
%

In the ``node'' pinning scheme (Fig.~\ref{fig:intel_MPI}~(a), black line), the  transition from communication free computing ($1$ node) to $2$ compute nodes with MPI communication brings a speedup of 20 \%. 
An approximate linear reduction of elapsed time is observed until $16$ nodes and from $>16$ nodes onwards a saturation behavior is observed. 

For the ``$2$-socket'' pinning scheme  (Fig.~\ref{fig:intel_MPI}~(a), red line) computational resources of nodes are exploited even more efficiently, i.e., elapsed time is reduced by a factor of $1.51$ on average compared to the ``node'' pinning scheme. 
Nevertheless,  numerical scaling with the number of nodes is reduced which is reflected in the ratio of elapsed times for $1$ and $20$ nodes, being  $1.55$ for the ``$2$-socket'' pinning scheme  and  $2.36$ for the "node" pinning scheme. 
%


The distribution of memory resources over multiple nodes is analyzed in Fig.~\ref{fig:intel_MPI}~(b)
for the "node"  and ``$2$-socket'' pinning schemes and compared to the ideal memory scaling (red dotted  line).
 For $1-4$ compute nodes, the ``node'' pinning scheme allows to use memory resources more efficiently than the ``$2$-socket'' scheme due to the extra copies of identical paths that can reside on different sockets but still within the same node.\footnote{Note that peak memory in the ``$2$-socket'' pinning scheme is an upper bound as peak memory is assessed on a per process and not on a per node basis. 
 In the case of two processes per node, we estimate peak memory usage by multiplying the process peak memory by two. However, it is highly improbable that the two most memory-consuming processes will reside on the same node.} 
For $> 4$ nodes, we find that paths are distributed more evenly over the larger number of MPI processes which reduces the number of duplicate copies residing on the same node. Thus the observed differences between peak, optimal, and minimum memory consumption per process are reduced. 
 Overall, we find a favorable real-to-ideal memory usage ratio that on average equals to $2.11$ ($> 2$ nodes), stressing the low memory overheads of the developed MPI implementation of inter-process path merging (cf. Section~\ref{sec:InterNode}).

\begin{figure*}
\includegraphics[width=0.49\textwidth]{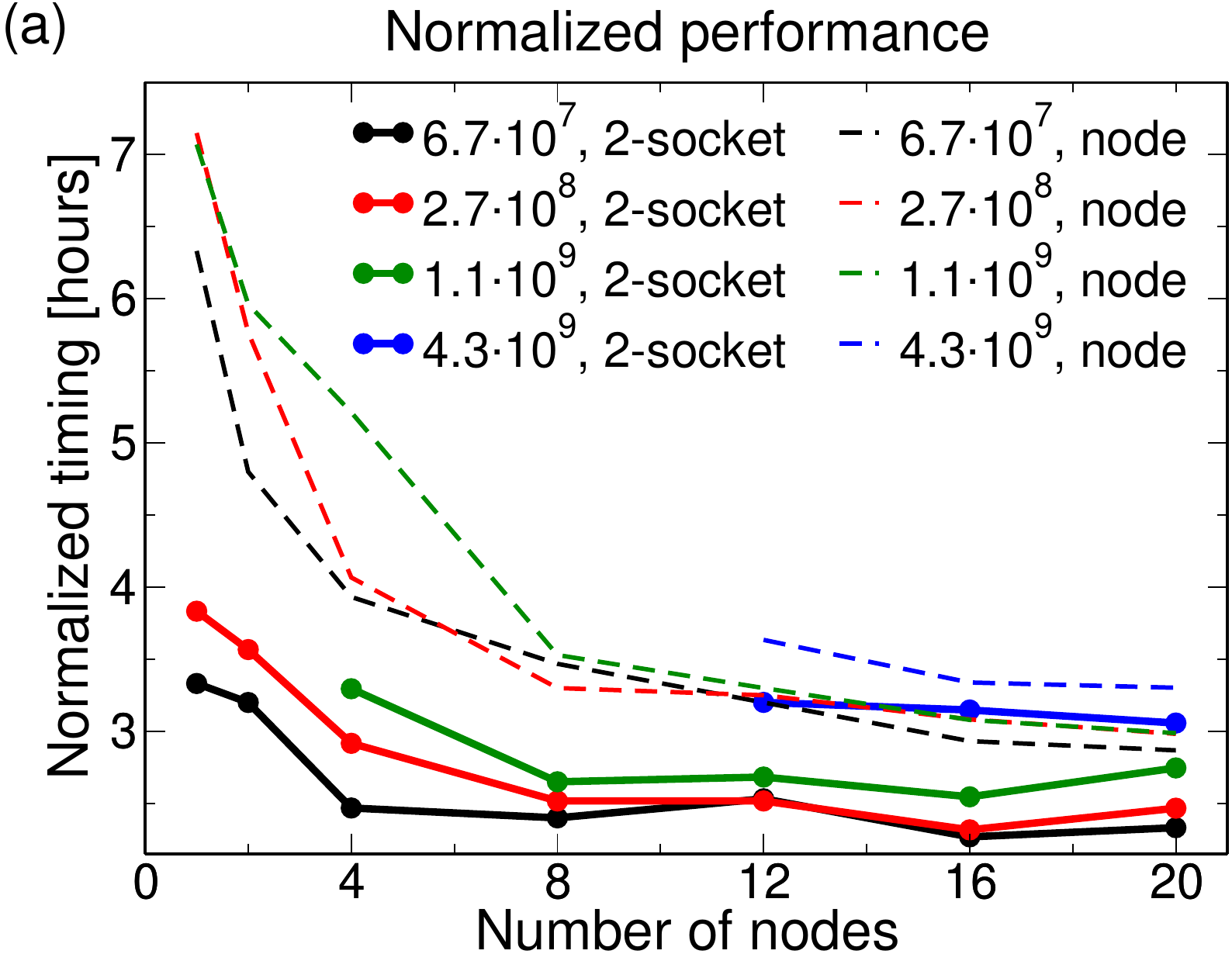}
\includegraphics[width=0.49\textwidth]{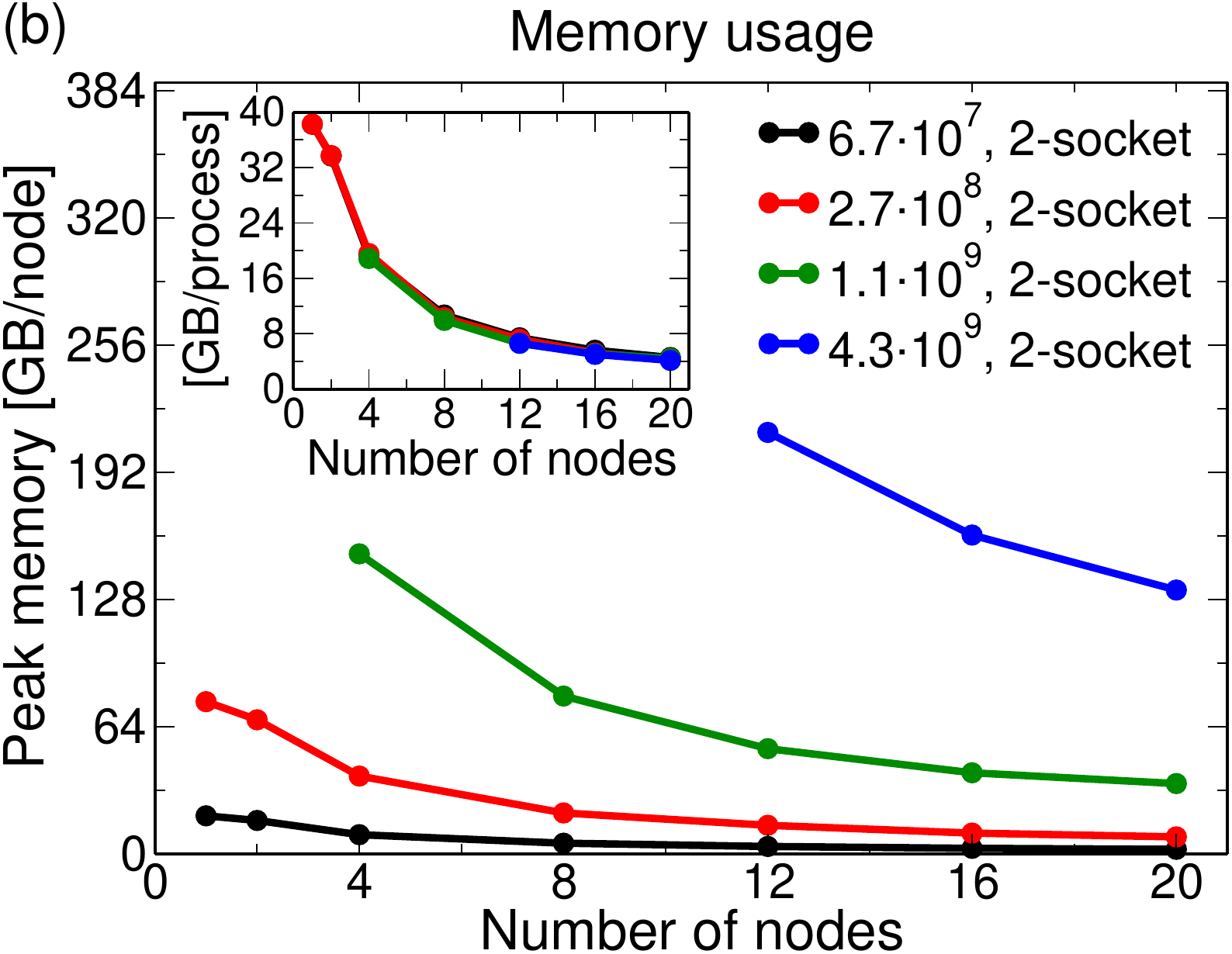}
    \caption{Numerical performance (a) and peak memory usage per computing node (b), all compute nodes are equipped with $256$~GB of memory. The inset of panel (b) gives the peak memory usage per MPI process.
    Numerical performance and peak memory usage per MPI process have been normalized to $2.7 \cdot 10^{8}$ paths, i.e., for $6.7 \cdot 10^7$, $2.7 \cdot 10^8$, $1.1 \cdot 10^9$, and $4.3 \cdot 10^9$ paths the normalization factors are $4$, $1$, $1/4$, and $1/16$, respectively. 
    The simulation model is the same as in Fig.~\ref{fig:intel_MPI}, but bath memory time is varied from $13$ time steps ($6.7 \cdot 10^7$ paths) to $16$ time points ($4.3 \cdot 10^9$ paths). 
    }
\label{fig:mpi_scaling}
\end{figure*}
%
Distributed memory parallelization allows to target higher problem complexity (reflected in the number of paths) by distributing  hardware requirements over multiple compute nodes. 
We thus investigate the scaling behavior of the MPI algorithm for an increasing number of paths by successively increasing  the bath memory time ($\Delta k_{\rm max} = \Delta k_{\rm eff} = 13$-$16$) which translates into $6.7\cdot10^7$ - $4.3\cdot10^9$ considered paths. 
Particular encouraging scaling behavior is observed for used peak memory (Fig.~\ref{fig:mpi_scaling}~(b)) that facilitates  the efficient exploitation of the computational resources of multiple nodes to overcome the memory bottleneck of QUAPI computations which is a key advantage of the distributed memory parallel algorithm.
Specifically, employed compute nodes are equipped with $256$~GB RAM which allows simulations with  $6.7 \cdot 10^7$ ($\Delta k_{\rm max} =13$) and $2.7 \cdot 10^8$ paths ($\Delta k_{\rm max} =14$) to be performed on a single compute node,
larger simulations are not possible on a single node due to memory constraints.
 Such computations required $\sim 34$~GB of memory per MPI process
 ($2.7 \cdot 10^8$ paths, $\sim 68$~GB per node, using $2$ computing nodes).

 In a numerically exact, full QUAPI simulation ($\Delta k_{\rm max} = \Delta k_{\rm eff}$, no filtering), increasing the memory time by a single time step leads to an increase of system complexity to  $1.1 \cdot 10^9$ paths ($\Delta k_{\rm max} =15$) that requires $34 \times 4 = 136$~GB per MPI process ($272$~GB per node).
 The distribution over four nodes facilitates such simulations requiring  $76$ GB memory per process (Fig.~\ref{fig:mpi_scaling}~(b), green line, $152$~GB per node).
 Similarly, the largest computations with $4.3\cdot10^9$ paths ($\Delta k_{\rm max} = 16$)  were facilitated with $n \geq 12$ compute nodes. 
 
 For increasing problem complexity, we find that simulations on $n$ nodes $(n>2)$ reduce the memory requirements by 
 $2.11/n$ times, compared to an identical computation performed an on single compute node. 
The inlay in Fig.~\ref{fig:mpi_scaling}~(b) shows the peak memory usage per MPI process, normalized with respect to the number of processes. As all normalized values essentially coincide, effective usage of memory resources is achieved  for  increasing problem complexity.

\subsection{Non-Markovian Dynamics Subject to a Structured Environment}
\label{sec:struc_env}


For a demonstration of the capabilities of the developed pre-merging distributed memory implementation of the MACGIC-QUAPI method, we simulated the dissipative dynamics of a two-level quantum system (TLS)
subject to the interaction with a structured 
environment. 
The considered spectral density $J(\omega)$ is characterized by a sharp resonance with characteristic frequency $\Omega$ augmented by an Ohmic background ($J(\omega) \propto \omega)$:
%
%
%
\begin{equation}
J(\omega) = \frac{2 \alpha \omega \Omega^4}{(\Omega^2 - \omega^2)^2 + (2 \pi \kappa \omega \Omega)^2} ,
\label{eq:sharpSD}
\end{equation}
where 
$2 \pi \kappa \Omega$ describes the width of the resonance.
Structured environments of the form of Eq.~(\ref{eq:sharpSD}) impose strong non-Markovianity in the dynamics and 
pose a persistent challenge to simulations of the real time dynamics of open-quantum systems~\cite{a:Fingerhut_19,Gibben:PhysRevResearch:2020}
The model is relevant for excitation energy transfer in biological light harvesting complexes where discrete 
nuclear modes can modulate the transfer dynamics~\cite{Chin:NatPhys:2013} or novel quantum technologies, like photonic cavities or superconducting quantum interference (dc-SQUID) devices~\cite{a:Grifoni_04}.

The discrete underdamped mode at characteristic frequency $\Omega$ induces long-time system-bath correlations that can strongly affect the system dynamics. In the time-domain, the narrow peak in the spectral density given by Eq.~(\ref{eq:sharpSD}) leads to a slowly oscillatory decay of the corresponding influence coefficients and thus a long-time non-Markovian memory that is challenging to converge in QUAPI simulations. 
Specifically,  for the investigated parameter regime $2 \pi \kappa \Omega$ $\approx$ $35$ cm$^{-1}$ (see below), the memory time spans some $\approx 228$ time steps (until influence coefficients drop to $1$\% of the zero-time lag value).



For numerical solution we resort to the a primary reaction coordinate representation~\cite{a:Ambegaokar_1985, a:Haenggi_00, Iles-Smith:PhysRevA:2014} where the TLS interacts with a primary mode that in turn is coupled to a
harmonic environment: 
an Ohmic environment:
%
\begin{equation}
%
\begin{aligned}
    H & = \frac{\hbar \Delta}{2} \sigma_x + \hbar g \sigma_z (B^\dagger + B) + \hbar\Omega B^\dagger B \\
    & + (B^\dagger + B) \sum_i \hbar \nu_i (b_i^\dagger + b_i) + \sum_i \hbar \omega_k b_i^\dagger b_i~.
\end{aligned}
\label{eq:tot_TSS_HO}
\end{equation}
The Hamiltonian of Eq.~(\ref{eq:tot_TSS_HO}) 
represents an effective quantum system with extended Hilbert space, composed of the electronic and vibrational degrees of freedom of the primary mode 
(first three terms in Eq.~(\ref{eq:tot_TSS_HO}), denoted $H_{TLS+HO}$).
We consider a bias-free TLS ($\epsilon = 0$), and 
$B$ and $B^\dagger$ denote annihilation and creation operators of the primary mode, respectively.
The operators $b_i$ and $b_i^\dagger$ are the corresponding annihilation and creation  operators of bath modes and $\sigma_{x/z}$ are  Pauli matrices. 

It has been demonstrated\cite{a:Grifoni_04} that the form of Eq.~(\ref{eq:tot_TSS_HO}), where via canonical transformation part of the  environment has been absorbed into the extended system $H_{TLS+HO}$, is more amenable for simulations with the QUAPI method.
The effective bath of $H_{TLS+HO}$, represented by oscillators $b_i$, has the form of an Ohmic spectral density $J(\omega) = \kappa \omega \, \EXP{-\omega / \omega_c}$ with system-bath coupling constant $\kappa$ and cut-off frequency $\omega_c$.
The original system-bath coupling $\alpha$, coupling constant $\kappa$ and the coupling $g$, describing the interaction between the TLS and the primary mode, are interconnected:
$\alpha \Omega^2 = 8 \kappa g^2$.
Generally, the effective  system-bath interaction $\kappa$ may be chosen small and the bath memory time is substantially reduced by the simplified form of $J(\omega)$.
In practice, the harmonic oscillator basis of the extended 
system $H_{TLS+HO}$ has to be truncated.

\begin{figure*}
\includegraphics[width=0.49\textwidth]{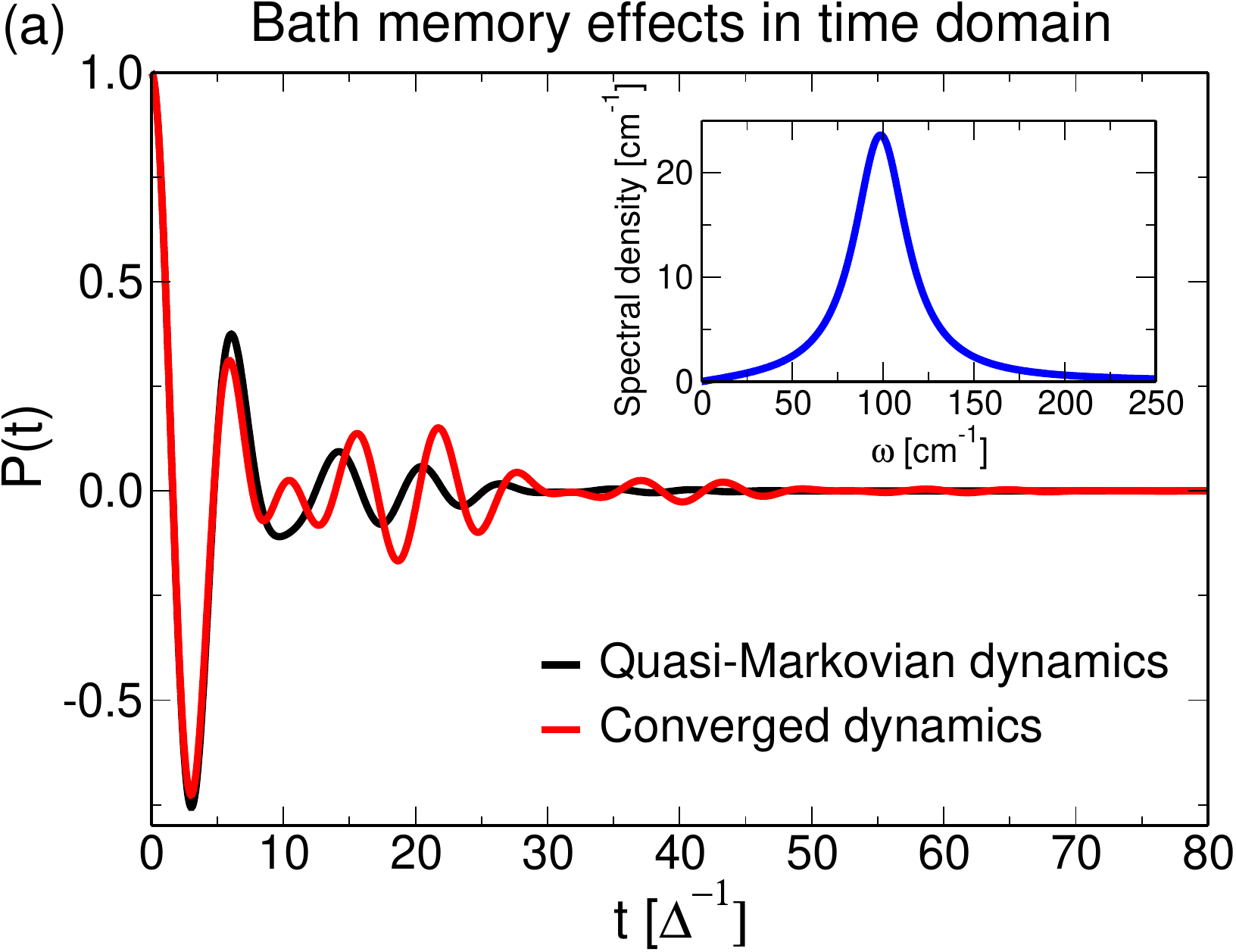}
\includegraphics[width=0.49\textwidth]{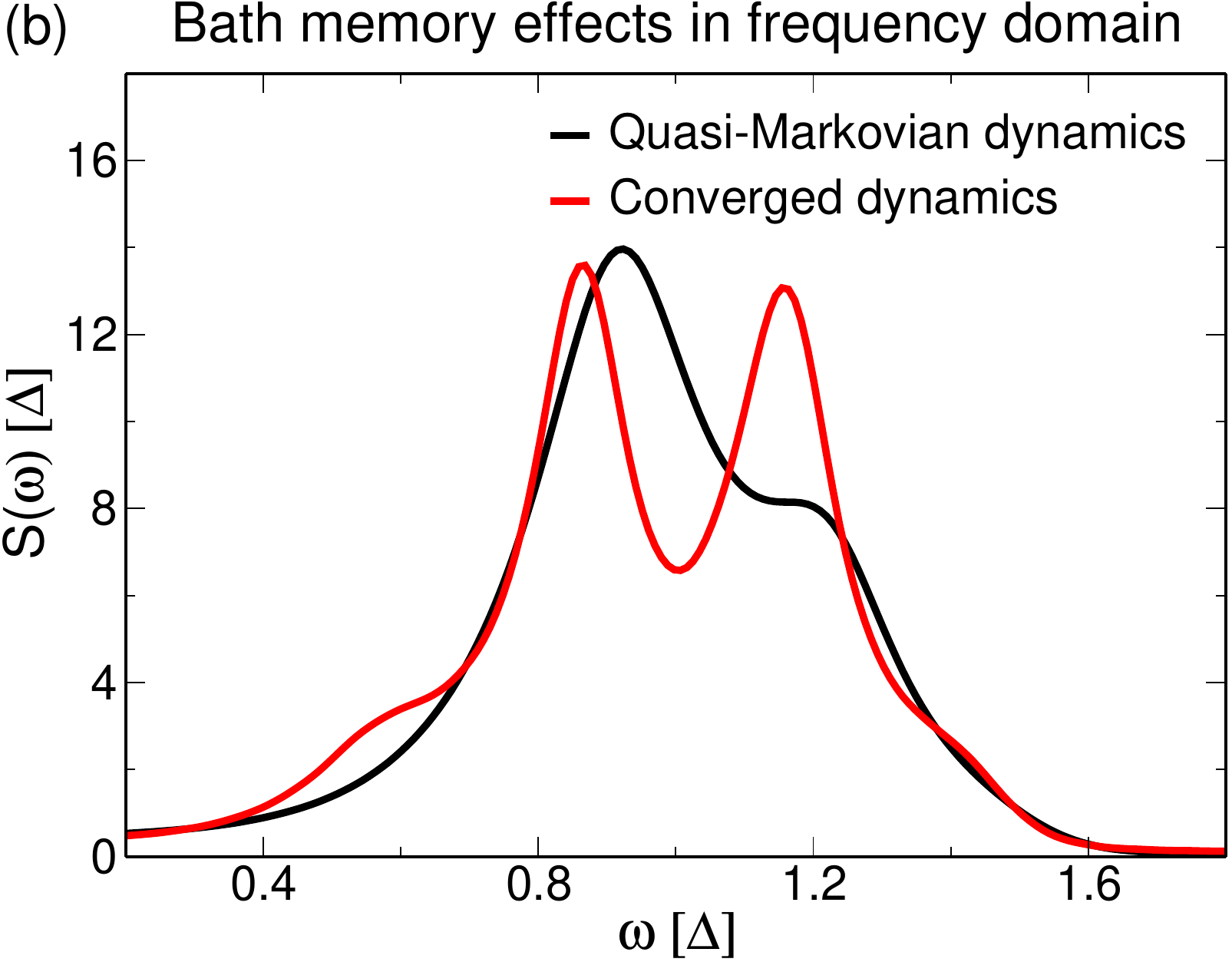}
    \caption{
    Converged dynamics (red) and quasi-Markovian dynamics (black) of the TLS in the time (a) and frequency domain $S(\omega)$ (b). 
    Shown is the electronic difference population, i.e., $P(t) = \langle \sigma_z \rangle (t)$ with the trace taken over vibrational states.
    The structured spectral density of the original model (Eq.~(\ref{eq:sharpSD})) is shown as inset in panel (a). Finite temperature is $k_BT = \hbar \Delta$ (with Boltzmann constant $k_B$) and  weak effective coupling between the primary mode $\Omega$ and the Ohmic environment ($\kappa = 0.056$), the coupling $g$ 
    is $g = 0.18 \Delta$. 
    Cut-off frequency $w_c = 10 \Delta$ and propagation time step  $\Delta t = 0.06 / \Delta$ (taken from Ref.~\citenum{a:Grifoni_04}) were confirmed to be sufficiently converged.
   $S(\omega)$ is obtained as $S(\omega) = 1/(2 \pi N_t) \cdot {\rm Re} \left\{ {\rm DFT} \left[ P(t) \right] \right\}$, where $N_t$ is the number of time sampling points and ${\rm DFT}$ denotes Discrete Fourier Transform. 
   }
\label{fig:conv_vs_markov}
\end{figure*}

Figure~\ref{fig:conv_vs_markov} (a) presents the real time dynamics of the TLS in resonance with the structured environment ($\Omega = \Delta$). The dynamics is solved numerically for the extended system $H_{TLS+HO}$ via Eq.~(\ref{eq:tot_TSS_HO}), 
simulations were performed at finite temperature $k_BT = \hbar \Delta$ (with Boltzmann constant $k_B$) and reasonably weak effective coupling between the primary mode and the Ohmic environment ($\kappa = 0.056$, simulation parameters are summarized in the caption of Fig.~\ref{fig:conv_vs_markov}). 
We investigated in detail the convergence properties with respect to memory time and truncation of the harmonic oscillator basis (cf. Sec.~\ref{sec:convergence}) and compare in Fig.~\ref{fig:conv_vs_markov} (a) the converged dynamics 
and the quasi-Markovian dynamics 
where system-bath correlations have been truncated at $\Delta k_{\rm max}=2$.
Evidently, even for the weak $\kappa = 0.056$, substantial differences are apparent: 
the converged dynamics appears highly non-Markovian and complex, while the quasi-Markovian dynamical simulations 
only reproduce the short-time dynamics until  $t \approx 10 \Delta^{-1}$. 
Later on, both simulations substantially deviate 
due to the long-time non-Markovian system-environment correlations.
Thus,  even for weak system-environment interaction $\kappa$ 
the non-Markovian memory between the extended quantum system and the environment has a strong impact on the dynamics.

%

%
%

The non-Markovian memory of the Ohmic environment has a profound impact on the dynamics, reflected in a re-emergence of coherence  at about $t \approx 15-25 \Delta^{-1}$, re-normalization of the oscillation frequency (see below) and prolongation of the coherence timescale until $t \approx 40-50 \Delta^{-1}$, reminiscent of the  long-lasting electronic coherence induced by  non-trivial spectral structures~\cite{Chin:NatPhys:2013}. 
%
%
In the  frequency domain ($S(\omega)$, Fig.~\ref{fig:conv_vs_markov} (b)), the converged dynamics is characterized by 
a splitting of peaks around the resonance frequency $\Omega = \Delta$ with peak position at $\omega = 0.87 \Delta$  and  $\omega = 1.15 \Delta$ with similar amplitudes due to the hybridization of electronic system and primary vibrational mode $\Omega$.
Comparing the quasi-Markovian dynamics, neither peak splitting, peak positions nor amplitude are reproduced.

The observed resonances in $S(\omega)$ (Fig.~\ref{fig:conv_vs_markov} (b)) correspond to transitions between the ground state $\KET{0; 0}$ ($E_0$) and the first two excited states $\{\KET{1; 0}; \KET{0; 1}\}$ 
of the extended system $H_{TLS+HO}$, inducing the splitting into $E_1$ and $E_2$. 
Here we adopt a notation, where the first  index in the parentheses denotes the occupation of the pure TLS Hamiltonian and the second index of the  pure HO Hamiltonian (see SI for the list of $H_{TLS+HO}$ Hamiltonian eigenvalues).
The corresponding transition energies are $E_1 - E_0 = 0.82 \Delta$ and $E_2 - E_0 = 1.18 \Delta$, in good agreement with the observed resonances.
Moreover, low- and high-frequency shoulders ($\omega = 0.55 \Delta$, and $1.42 \Delta$) arise in $S(\omega)$ of the converged dynamics. Such transitions can be assigned to the eigenstate differences $E_3 - E_2 = 0.57 \Delta$ and $E_4 - E_1 = 1.43 \Delta$ of the extended $H_{TLS+HO}$ Hamiltonian with predominant character 
$\{\KET{0; 2}; \KET{1; 1}\}$ for $E_3$ and $E_4$, and $\{\KET{0; 1}; \KET{1; 0}\}$  for $E_2$ and $E_1$ in the original basis. 
%
For the close-to Markovian case,
the side-bands are absent and the high energy resonance (corresponding to $E_2 - E_0$) is not fully resolved.
Similarly, the low energy peak position ($\omega = 0.87 \Delta$) significantly deviates from the reference value and teh intensity pattern is not captured.
The higher spectral resolution in the converged simulation is a consequence of the longer lived coherence and reduced dephasing rate (estimated as FWHM of the corresponding peaks) 
while 
the total FWHM of the spectrum is reasonably captured.

We would like to note, that the presented strongly non-Markovian simulations in the primary mode representation of Eq.~(\ref{eq:tot_TSS_HO}) are computationally demanding. 
 For the converged dynamics shown in Fig.~\ref{fig:conv_vs_markov} runtime was $\sim 32$~hours on the $96$ cores of a four socket Intel Xeon Platinum 8160 processor system (2 TB shared memory).
 The simulations considered on average $2.5 \cdot 10^8$ paths (mask size $\Delta k_{\mathrm{eff}} = 10$, filter threshold  $\theta = 10^{-9}$, leading to  a loss of $\sim 0.2 \%$ in the normalization of the reduced density matrix at the end of propagation.
 The peak memory requirement was $24$ - $26$~GB per MPI process or at most $104$~GB per node (taking into account duplicating data due to the inter-process mask merging).
 

\subsubsection{Parameter Convergence:}
\label{sec:convergence}

\begin{figure*}[t!]
\includegraphics[width=0.49\textwidth]{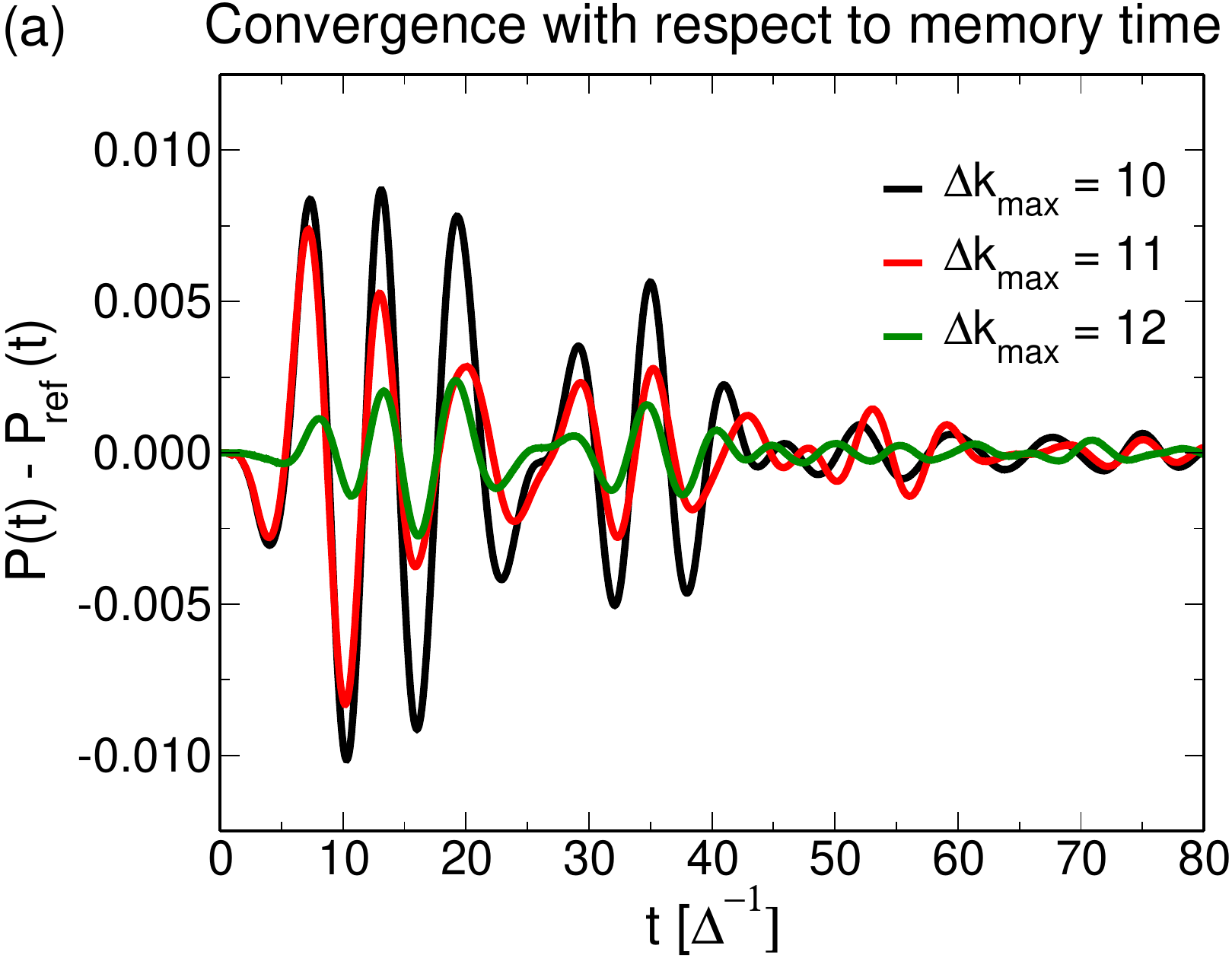}
\includegraphics[width=0.49\textwidth]{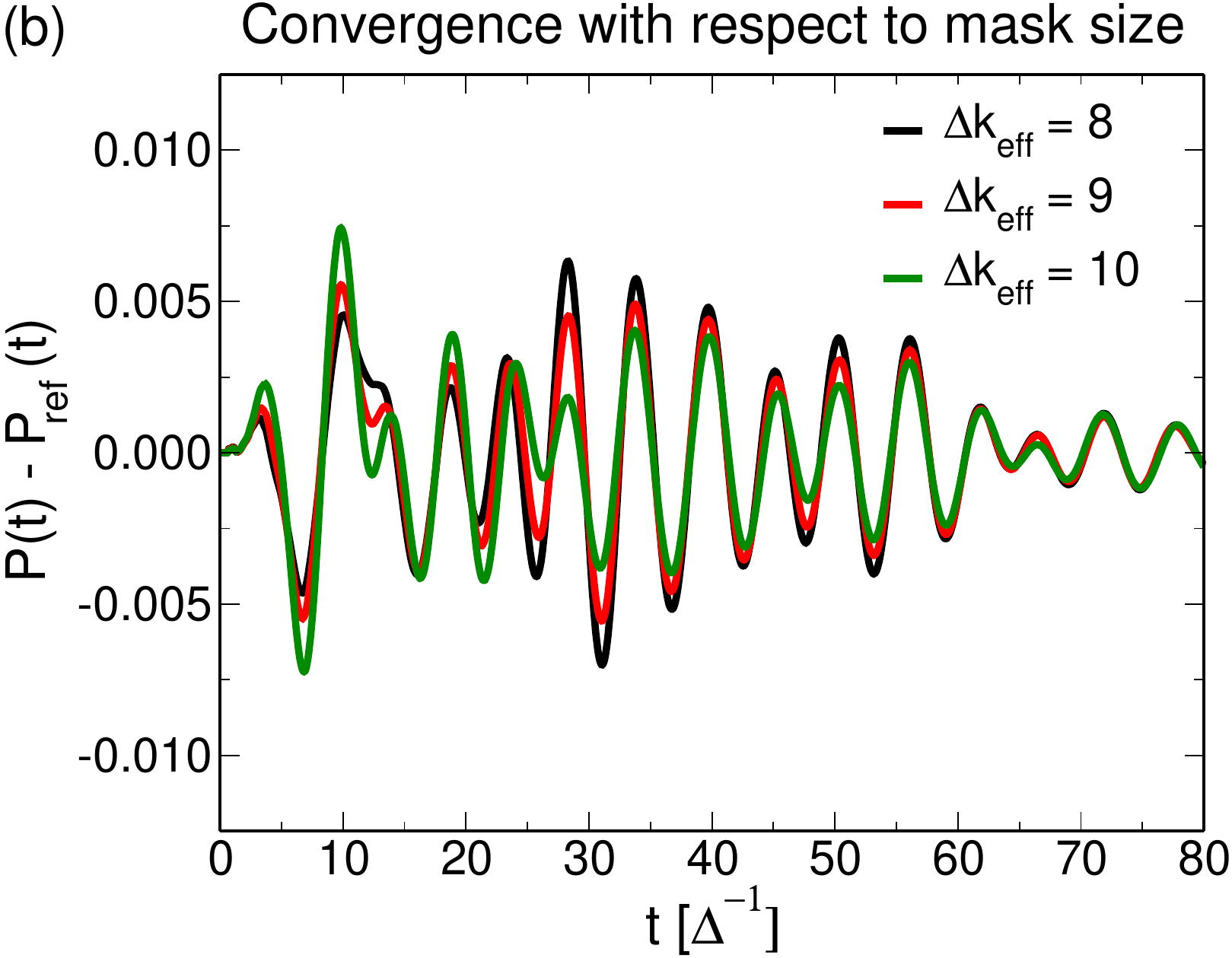}
\\
\includegraphics[width=0.49\textwidth]{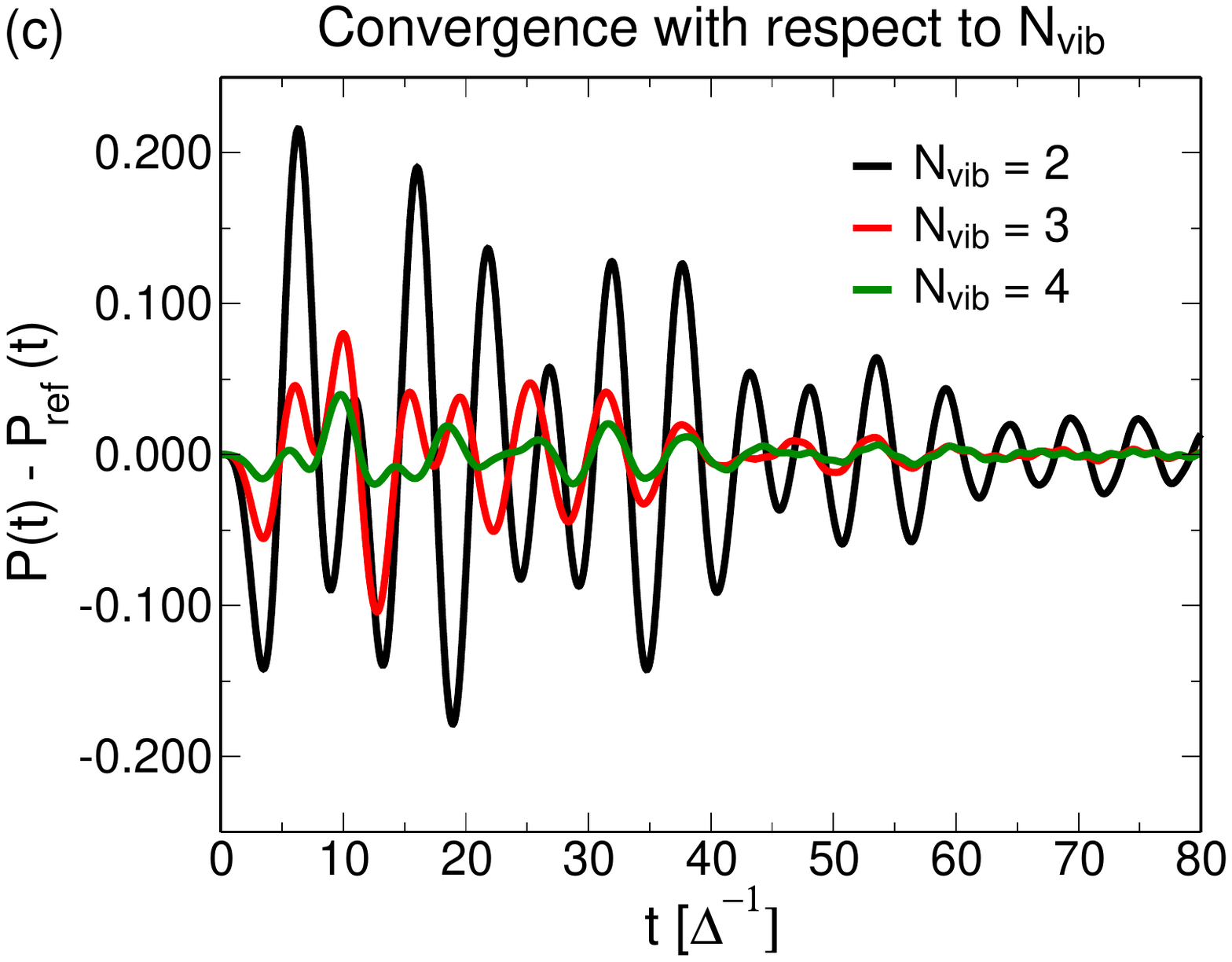}
\includegraphics[width=0.49\textwidth]{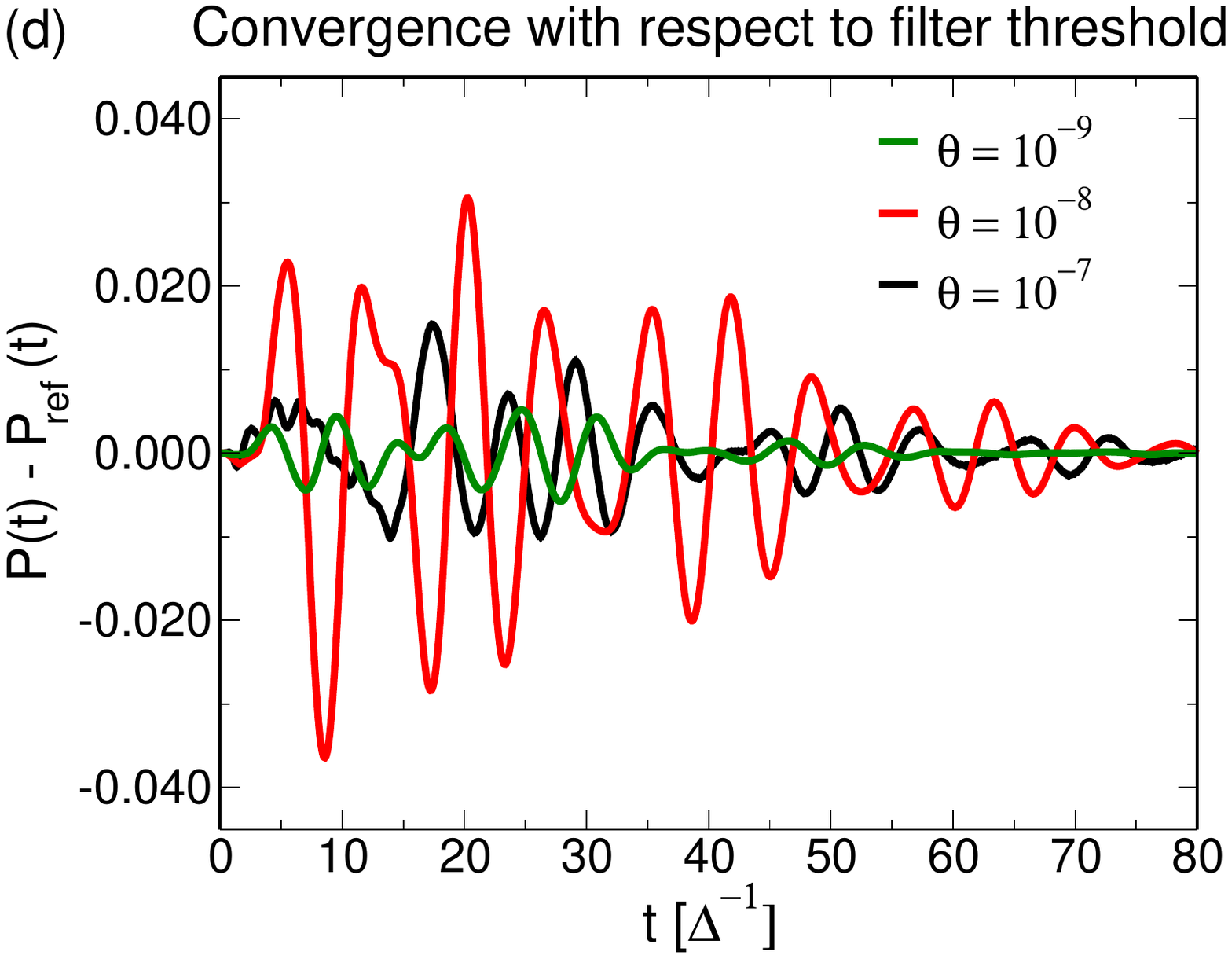}
\caption{Parameter convergence with respect to bath memory time $\Delta k_{\mathrm{max}}$ (a), mask size $\Delta k_{\mathrm{eff}}$ (b), harmonic oscillator basis size $N_{\mathrm{vib}}$ (c) and filter threshold $\theta$ (d). 
The difference of population is shown for each trial parameter with respect to a reference setting ($P_{\mathrm{ref}}(t)$). 
Reference setting is $\Delta k_{\rm max} = 15$ in panel (a),
 $\Delta k_{\rm eff} = 11$ in panel (b), 
 $N_{\rm vib} = 5$ in panel (c) and $\theta = 10^{-11}$ in (d).
 The remaining parameters in reference calculations of $P_{\mathrm{ref}}(t)$ are fixed at their converged values, i.e., $\Delta k_{\rm max} = 12$, $\Delta k_{\rm eff} = 10$, $N_{\rm vib} = 4$, and $\theta = 10^{-9}$; see Table~S3 for a detailed summary all simulation parameter settings.}
\label{fig:dyn_pt_set_conv}
\end{figure*}

Figure~\ref{fig:dyn_pt_set_conv} demonstrates the convergence of simulation parameters, i.e.,  memory time $\Delta k_{\rm max}$, mask size $\Delta k_{\rm eff}$, truncation level of the harmonic oscillator basis of the primary mode $N_{\rm vib}$ and path filtering threshold $\theta$.
For  $\Delta k_{\rm max}=10$, the absolute error in population is $\approx 0.01$ which can be systematically reduced  to  $< 0.0025$ for  $\Delta k_{\rm max}=12$. Note that due to the MAGCIC approximation, the increase in total bath memory time does not determine the numerical effort of simulations and extensive simulations up to $\Delta k_{\rm max} = 15$ have been performed.
Nevertheless, long-time system bath correlation with $\Delta k_{\rm max}>10$, i.e., the timescale of the primary mode,  is required to accurately describe the dynamics even for relatively weak coupling to the Ohmic environment.
For the convergence of mask size (Fig.~\ref{fig:dyn_pt_set_conv} (b)),  $\Delta k_{\rm eff}=10$ allows for converged dynamics on the order  $\approx 0.0075$. We note slightly irregular convergence behavior with increasing $\Delta k_{\rm eff}$ where short time differences increase ($t < 25 \Delta^{-1}$) while the long time differences ($t > 25 \Delta^{-1}$) decrease.
In contrast, the filtering threshold shows  monotonic convergence and converged dynamics ($\approx 0.006$) is achieved with $\theta$ = $10^{-9}$ (Fig.~\ref{fig:dyn_pt_set_conv} (d)).

The truncation level of the harmonic oscillator basis (Fig.~\ref{fig:dyn_pt_set_conv} (c)) shows the most pronounced effect onthe dynamics.
Needless to say, the accompanying increase in the size of the Hilbert space ($M = 2 N_{\rm vib}$) also poses the greatest challenges for investigations of the numerical convergence.
As the simulations are performed at finite temperature ($k_BT = \hbar \Delta$) thermal excitation of vibrational excited states affects the system dynamics. We find that converged dynamics (error $\approx 0.04$) can be achieved with $N_{\rm vib} = 4$.

\subsubsection{Positive and Negative Detuning $\Omega \neq \Delta$:}
%
\begin{figure*}
\includegraphics[width=0.49\textwidth]{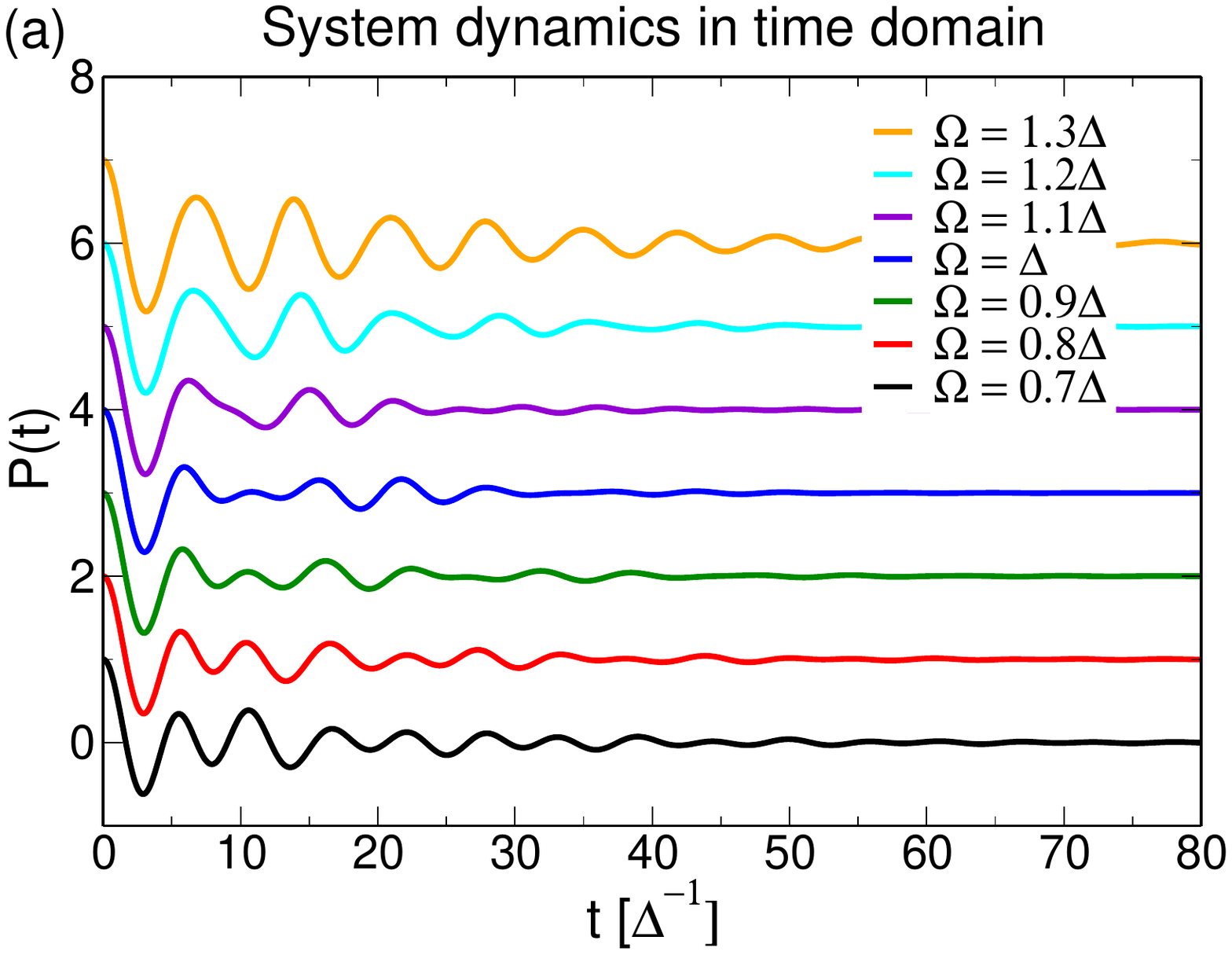}
\includegraphics[width=0.49\textwidth]{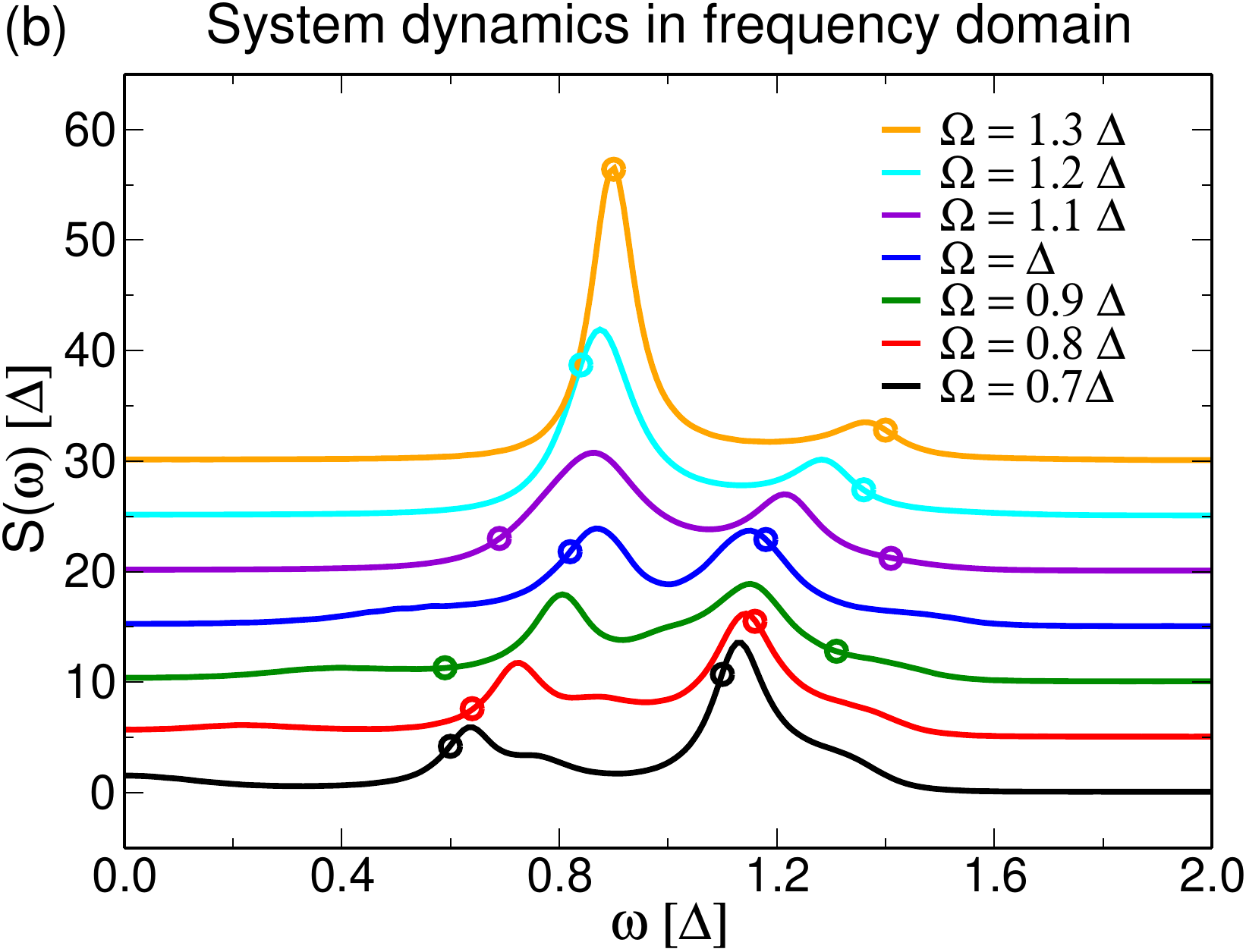}
    \caption{System dynamics in the time (a) and frequency (b) domains for varying HO transition frequency $\Omega$. 
    Simulations are performed for the Hamiltonian of Eq.~(\ref{eq:tot_TSS_HO}), parameter settings are: $k_B T = \hbar \Delta$, $\kappa = 0.056$, $g = 0.18 \Delta$, and $\omega_c = 10 \Delta$. The QUAPI parameters for converged dynamics are: $\Delta t = 0.06 \Delta^{-1}$, $\Delta k_{\rm max} = 12$, $\Delta k_{\rm eff} = 10$, $N_{\rm vib} = 4$, and $\theta = 10^{-9}$. Peak positions according to first order perturbation theory with respect to the interaction parameter $g$ are shown as circles in panel (b). The detailed list of eigenvalues for the total Hamiltonian is presented in the SI.}
\label{fig:dyn_pt_set_omega}
\end{figure*}
Figure~\ref{fig:dyn_pt_set_omega} presents the converged real time dynamics for the tuning of the primary harmonic oscillator frequency $\Omega$.
For large positive detuning ($\Omega = 1.3 \Delta$) the dynamics is dominated by a single oscillation frequency with weak modulations only. Accordingly, in the frequency domain the spectrum shows a dominant peak at $\omega = 0.90 \Delta$ and a weaker peak at $\omega = 1.36 \Delta$.  
Decreasing the detuning leads to more complex dynamics in the time domain which is reflected in comparable spectral amplitudes of peaks in the spectra.   
Around the resonance condition ($1.1 \Delta \geq \Omega \geq 0.9 \Delta$),  dephasing becomes maximal which is reflected in a broadening of peaks in Fig.~\ref{fig:dyn_pt_set_omega} (b). 
Accelerated dephasing arises from the strong mixing of vibronic states, the
electronic degrees of freedom are thus most directly affected by the Ohmic bath in this regime. For negative detuning  ($\Omega < 0.9 \Delta$) most complex real time dynamics arises which is reflected in the emerging shoulders in the spectrum due to  vibrational excitations.
 
Perturbation theory predicts frequency positions $\Omega-\frac{g^2}{\Delta-\Omega}$ and $\Delta+\frac{g^2}{\Delta-\Omega}$ for the low  and high energy  peaks, respectively.
At resonance $\Omega =  \Delta$, both peaks have the same intensity and should be located at $\Delta-g = 0.82 \Delta$ and $\Delta+g = 1.18 \Delta$ which is in good agreement with the numerical values of $0.87 \Delta$ and $1.15 \Delta$, respectively. We assign the differences to a non-trivial renormalization of system resonance frequency due to the non-perturbative system bath interaction already at $\kappa = 0.056$. While for positive and negative detuning the frequency position of the weaker peak reasonably follows the perturbative prediction, substantial deviations in frequency position of the high intensity peak are recognized. Arising sideband shoulders  due to vibrational excitations among excited states are not accounted for by the perturbative treatment

\section{Conclusions}

We have presented a new path pre-merging algorithm of the MACGIC-QUAPI method that 
mitigates the memory bottleneck of the method while preserving numerical accuracy.
The used hash map data structure facilitates dynamical memory allocation and the
efficient look-up paths, and is well suited for 
%
the irregular allocation requirements of the mask merging algorithm with generally sparse masks.



Performance increases are realized via a hybrid MPI-Intel TBB parallel implementation of the algorithm.  The scalable distributed memory parallelization allows to spread the considered paths among  memory of multiple compute nodes, thereby overcoming the memory bottleneck inherent to QUAPI methods. We showed that the current implementation provides  a memory scaling as $2.11/n$ with $n$ being the number of nodes. Thus, the peak memory usage per node may be substantially reduced in computations.
Scaling of the algorithm has been explored up to $n=20$ nodes, making the implementation suitable for supercomputing architectures.


The distributed memory implementation facilitates large scale numerical simulations and was applied to investigate the non-Markovian dynamics imposed by a structured environment. 
 The numerical challenging problem of ultra-long system bath correlation times is recast by including the primary mode explicitly into the system Hilbert space which makes the problem tractable for MACGIG-QUAPI simulations and facilitates the investigation of convergence properties.
We find that even for weak coupling of the extended system to an effective Ohmic bath, the system dynamics is highly non-Markovian.
The detuning of the the primary mode frequency demonstrates the limits of perturbative treatments, as reflected in pronounced sidebands due to vibrational excitations. 

\begin{acknowledgement}
This research has received funding from the European Research Council under the European Unions Horizon 2020 and Horizon Europe Research and Innovation programs (Grant Agreements No. 802817 and  No. 101125590).
The authors would like to thank the Deutsche Forschungsgemeinschaft for ﬁnancial support [Cluster of Excellence \textit{e}-conversion (Grant No. EXC2089-390776260)].
The authors acknowledge the computational and data resources provided by the Leibniz Supercomputing Centre (www.lrz.de).

\end{acknowledgement}

\bibliography{biblio}

\providecommand{\latin}[1]{#1}
\makeatletter
\providecommand{\doi}
  {\begingroup\let\do\@makeother\dospecials
  \catcode`\{=1 \catcode`\}=2 \doi@aux}
\providecommand{\doi@aux}[1]{\endgroup\texttt{#1}}
\makeatother
\providecommand*\mcitethebibliography{\thebibliography}
\csname @ifundefined\endcsname{endmcitethebibliography}
  {\let\endmcitethebibliography\endthebibliography}{}
\begin{mcitethebibliography}{46}
\providecommand*\natexlab[1]{#1}
\providecommand*\mciteSetBstSublistMode[1]{}
\providecommand*\mciteSetBstMaxWidthForm[2]{}
\providecommand*\mciteBstWouldAddEndPuncttrue
  {\def\EndOfBibitem{\unskip.}}
\providecommand*\mciteBstWouldAddEndPunctfalse
  {\let\EndOfBibitem\relax}
\providecommand*\mciteSetBstMidEndSepPunct[3]{}
\providecommand*\mciteSetBstSublistLabelBeginEnd[3]{}
\providecommand*\EndOfBibitem{}
\mciteSetBstSublistMode{f}
\mciteSetBstMaxWidthForm{subitem}{(\alph{mcitesubitemcount})}
\mciteSetBstSublistLabelBeginEnd
  {\mcitemaxwidthsubitemform\space}
  {\relax}
  {\relax}

\bibitem[Wei\ss(2008)]{b:Weiss_08}
Wei\ss,~U. \emph{Quantum Dissipative Systems}; World Scientific, Singapore,
  2008\relax
\mciteBstWouldAddEndPuncttrue
\mciteSetBstMidEndSepPunct{\mcitedefaultmidpunct}
{\mcitedefaultendpunct}{\mcitedefaultseppunct}\relax
\EndOfBibitem
\bibitem[Breuer and Petruccione(2002)Breuer, and Petruccione]{b:Petruccione_02}
Breuer,~H.-P.; Petruccione,~F. \emph{The Theory of Open Quantum Systems};
  Oxford University Press: Oxford, 2002\relax
\mciteBstWouldAddEndPuncttrue
\mciteSetBstMidEndSepPunct{\mcitedefaultmidpunct}
{\mcitedefaultendpunct}{\mcitedefaultseppunct}\relax
\EndOfBibitem
\bibitem[May and K\"uhn(2011)May, and K\"uhn]{b:Kuehn_11}
May,~V.; K\"uhn,~O. \emph{Charge and Energy Transfer Dynamics in Molecular
  Systems}; Wiley-VCH, Weinheim, Germany, 2011\relax
\mciteBstWouldAddEndPuncttrue
\mciteSetBstMidEndSepPunct{\mcitedefaultmidpunct}
{\mcitedefaultendpunct}{\mcitedefaultseppunct}\relax
\EndOfBibitem
\bibitem[Makhlin \latin{et~al.}(2001)Makhlin, Schön, and
  Shnirman]{a:Shnirman_01}
Makhlin,~Y.; Schön,~G.; Shnirman,~A. Quantum-state engineering with
  Josephson-junction devices. \emph{Rev.~Mod.~Phys.} \textbf{2001}, \emph{73},
  357\relax
\mciteBstWouldAddEndPuncttrue
\mciteSetBstMidEndSepPunct{\mcitedefaultmidpunct}
{\mcitedefaultendpunct}{\mcitedefaultseppunct}\relax
\EndOfBibitem
\bibitem[Van~der Wal \latin{et~al.}(2000)Van~der Wal, Ter~Haar, Wilhelm,
  Schouten, Harmans, Orlando, Lloyd, and Mooij]{a:Wal_00}
Van~der Wal,~C.~H.; Ter~Haar,~A. C.~J.; Wilhelm,~F.~K.; Schouten,~R.~N.;
  Harmans,~C. J. P.~M.; Orlando,~T.~P.; Lloyd,~S.; Mooij,~J.~E. Quantum
  Superposition of Macroscopic Persistent-Current States. \emph{Science}
  \textbf{2000}, \emph{290}, 773\relax
\mciteBstWouldAddEndPuncttrue
\mciteSetBstMidEndSepPunct{\mcitedefaultmidpunct}
{\mcitedefaultendpunct}{\mcitedefaultseppunct}\relax
\EndOfBibitem
\bibitem[Chiorescu \latin{et~al.}(2003)Chiorescu, Nakamura, Harmans, and
  Mooij]{a:Mooij_03}
Chiorescu,~I.; Nakamura,~Y.; Harmans,~K.; Mooij,~J. Coherent quantum dynamics
  of a superconducting flux qubit. \emph{Science} \textbf{2003}, \emph{299},
  1869\relax
\mciteBstWouldAddEndPuncttrue
\mciteSetBstMidEndSepPunct{\mcitedefaultmidpunct}
{\mcitedefaultendpunct}{\mcitedefaultseppunct}\relax
\EndOfBibitem
\bibitem[Jortner and Bixon(1999)Jortner, and Bixon]{Jortner:AdvChem:1999}
Jortner,~J., Bixon,~M., Eds. \emph{Adv. Chem. Phys.}; John Wiley \& Sons,
  Incorporated: New York, 1999\relax
\mciteBstWouldAddEndPuncttrue
\mciteSetBstMidEndSepPunct{\mcitedefaultmidpunct}
{\mcitedefaultendpunct}{\mcitedefaultseppunct}\relax
\EndOfBibitem
\bibitem[Tamura \latin{et~al.}(2011)Tamura, Burghardt, and
  Tsukada]{Tamura:JPCC:2011}
Tamura,~H.; Burghardt,~I.; Tsukada,~M. Exciton Dissociation at
  Thiophene/Fullerene Interfaces: The Electronic Structures and Quantum
  Dynamics. \emph{J. Phys. Chem. C} \textbf{2011}, \emph{115},
  10205--10210\relax
\mciteBstWouldAddEndPuncttrue
\mciteSetBstMidEndSepPunct{\mcitedefaultmidpunct}
{\mcitedefaultendpunct}{\mcitedefaultseppunct}\relax
\EndOfBibitem
\bibitem[de~Vega and Alonso(2017)de~Vega, and Alonso]{Vega:RevModPhys:2017}
de~Vega,~I.; Alonso,~D. Dynamics of non-Markovian open quantum systems.
  \emph{Rev. Mod. Phys.} \textbf{2017}, \emph{89}, 015001\relax
\mciteBstWouldAddEndPuncttrue
\mciteSetBstMidEndSepPunct{\mcitedefaultmidpunct}
{\mcitedefaultendpunct}{\mcitedefaultseppunct}\relax
\EndOfBibitem
\bibitem[Ishizaki and Fleming(2009)Ishizaki, and Fleming]{Ishizaki:JCP:2009}
Ishizaki,~A.; Fleming,~G.~R. On the adequacy of the Redfield equation and
  related approaches to the study of quantum dynamics in electronic energy
  transfer. \emph{J. Chem. Phys.} \textbf{2009}, \emph{130}, 234110\relax
\mciteBstWouldAddEndPuncttrue
\mciteSetBstMidEndSepPunct{\mcitedefaultmidpunct}
{\mcitedefaultendpunct}{\mcitedefaultseppunct}\relax
\EndOfBibitem
\bibitem[Nalbach and Thorwart(2010)Nalbach, and Thorwart]{Nalbach:JCP:2010}
Nalbach,~P.; Thorwart,~M. Multiphonon transitions in the biomolecular energy
  transfer dynamics. \emph{J. Chem. Phys.} \textbf{2010}, \emph{132},
  194111\relax
\mciteBstWouldAddEndPuncttrue
\mciteSetBstMidEndSepPunct{\mcitedefaultmidpunct}
{\mcitedefaultendpunct}{\mcitedefaultseppunct}\relax
\EndOfBibitem
\bibitem[Rosenbach \latin{et~al.}(2016)Rosenbach, Cerrillo, Huelga, Cao, and
  Plenio]{Rosenbach:NJP:2016}
Rosenbach,~R.; Cerrillo,~J.; Huelga,~S.~F.; Cao,~J.; Plenio,~M.~B. Efficient
  simulation of non-Markovian system-environment interaction. \emph{New J.
  Phys.} \textbf{2016}, \emph{18}, 023035\relax
\mciteBstWouldAddEndPuncttrue
\mciteSetBstMidEndSepPunct{\mcitedefaultmidpunct}
{\mcitedefaultendpunct}{\mcitedefaultseppunct}\relax
\EndOfBibitem
\bibitem[Iles-Smith \latin{et~al.}(2016)Iles-Smith, Dijkstra, Lambert, and
  Nazir]{Iles-Smith:JCP:2016}
Iles-Smith,~J.; Dijkstra,~A.~G.; Lambert,~N.; Nazir,~A. Energy transfer in
  structured and unstructured environments: Master equations beyond the
  Born-Markov approximations. \emph{J. Chem. Phys.} \textbf{2016}, \emph{144},
  044110\relax
\mciteBstWouldAddEndPuncttrue
\mciteSetBstMidEndSepPunct{\mcitedefaultmidpunct}
{\mcitedefaultendpunct}{\mcitedefaultseppunct}\relax
\EndOfBibitem
\bibitem[Gribben \latin{et~al.}(2020)Gribben, Strathearn, Iles-Smith, Kilda,
  Nazir, Lovett, and Kirton]{Gibben:PhysRevResearch:2020}
Gribben,~D.; Strathearn,~A.; Iles-Smith,~J.; Kilda,~D.; Nazir,~A.;
  Lovett,~B.~W.; Kirton,~P. Exact quantum dynamics in structured environments.
  \emph{Phys. Rev. Research} \textbf{2020}, \emph{2}, 013265\relax
\mciteBstWouldAddEndPuncttrue
\mciteSetBstMidEndSepPunct{\mcitedefaultmidpunct}
{\mcitedefaultendpunct}{\mcitedefaultseppunct}\relax
\EndOfBibitem
\bibitem[Feynmann(1948)]{a:Feynmann_48}
Feynmann,~R.~P. Space-Time Approach to Non-Relativistic Quantum Mechanics.
  \emph{Rev.~Mod.~Phys.} \textbf{1948}, \emph{20}, 367\relax
\mciteBstWouldAddEndPuncttrue
\mciteSetBstMidEndSepPunct{\mcitedefaultmidpunct}
{\mcitedefaultendpunct}{\mcitedefaultseppunct}\relax
\EndOfBibitem
\bibitem[Feynmann and Vernon(1963)Feynmann, and Vernon]{a:Vernon_63}
Feynmann,~R.~P.; Vernon,~J.~F.~L. The Theory of a general quantum system
  interacting with a linear dissipative system. \emph{Ann.~Phys.}
  \textbf{1963}, \emph{24}, 118\relax
\mciteBstWouldAddEndPuncttrue
\mciteSetBstMidEndSepPunct{\mcitedefaultmidpunct}
{\mcitedefaultendpunct}{\mcitedefaultseppunct}\relax
\EndOfBibitem
\bibitem[Tanimura and Kubo(1989)Tanimura, and Kubo]{a:Kubo_89}
Tanimura,~Y.; Kubo,~R. Time Evolution of a Quantum System in Contact with a
  Nearly Gaussian-Markoffian Noise Bath. \emph{J.~Phys.~Soc.~Jpn.}
  \textbf{1989}, \emph{58}, 101\relax
\mciteBstWouldAddEndPuncttrue
\mciteSetBstMidEndSepPunct{\mcitedefaultmidpunct}
{\mcitedefaultendpunct}{\mcitedefaultseppunct}\relax
\EndOfBibitem
\bibitem[Tanimura(1990)]{a:Tanimura_90}
Tanimura,~Y. Nonperturbative expansion method for a quantum system coupled to a
  harmonic-oscillator bath. \emph{Phys.~Rev.~A} \textbf{1990}, \emph{41},
  6676\relax
\mciteBstWouldAddEndPuncttrue
\mciteSetBstMidEndSepPunct{\mcitedefaultmidpunct}
{\mcitedefaultendpunct}{\mcitedefaultseppunct}\relax
\EndOfBibitem
\bibitem[Tanimura(2020)]{Tanimura:JCP:2020}
Tanimura,~Y. Numerically ``exact''approach to open quantum dynamics: The
  hierarchical equations of motion (HEOM). \emph{J. Chem. Phys.} \textbf{2020},
  \emph{153}, 020901\relax
\mciteBstWouldAddEndPuncttrue
\mciteSetBstMidEndSepPunct{\mcitedefaultmidpunct}
{\mcitedefaultendpunct}{\mcitedefaultseppunct}\relax
\EndOfBibitem
\bibitem[Makri and Makarov(1995)Makri, and Makarov]{a:Makarov_th_95}
Makri,~N.; Makarov,~D.~E. Tensor propagator for iterative quantum time
  evolution of reduced density matrices. I. Theory. \emph{J.~Chem.~Phys.}
  \textbf{1995}, \emph{102}, 4600\relax
\mciteBstWouldAddEndPuncttrue
\mciteSetBstMidEndSepPunct{\mcitedefaultmidpunct}
{\mcitedefaultendpunct}{\mcitedefaultseppunct}\relax
\EndOfBibitem
\bibitem[Makri and Makarov(1995)Makri, and Makarov]{a:Makarov_res_95}
Makri,~N.; Makarov,~D.~E. Tensor propagator for iterative quantum time
  evolution of reduced density matrices. II. Numerical methodology.
  \emph{J.~Chem.~Phys.} \textbf{1995}, \emph{102}, 4611\relax
\mciteBstWouldAddEndPuncttrue
\mciteSetBstMidEndSepPunct{\mcitedefaultmidpunct}
{\mcitedefaultendpunct}{\mcitedefaultseppunct}\relax
\EndOfBibitem
\bibitem[Lambert and Makri(2012)Lambert, and Makri]{a:Lambert_12}
Lambert,~R.; Makri,~N. Memory propagator matrix for long-time dissipative
  charge transfer dynamics. \emph{Mol.~Phys.} \textbf{2012}, \emph{110},
  1967\relax
\mciteBstWouldAddEndPuncttrue
\mciteSetBstMidEndSepPunct{\mcitedefaultmidpunct}
{\mcitedefaultendpunct}{\mcitedefaultseppunct}\relax
\EndOfBibitem
\bibitem[Strathearn \latin{et~al.}(2018)Strathearn, Kirton, Kilda, Keeling, and
  Lovett]{Strathearn:NatCom:2018}
Strathearn,~A.; Kirton,~P.; Kilda,~D.; Keeling,~J.; Lovett,~B.~W. Efficient
  non-Markovian quantum dynamics using time-evolving matrix product operators.
  \emph{Nat. Commun.} \textbf{2018}, \emph{9}, 3322\relax
\mciteBstWouldAddEndPuncttrue
\mciteSetBstMidEndSepPunct{\mcitedefaultmidpunct}
{\mcitedefaultendpunct}{\mcitedefaultseppunct}\relax
\EndOfBibitem
\bibitem[Schollwöck(2011)]{Schollwock:AnnPhys:2011}
Schollwöck,~U. The density-matrix renormalization group in the age of matrix
  product states. \emph{Ann. Phys.} \textbf{2011}, \emph{326}, 96--192, January
  2011 Special Issue\relax
\mciteBstWouldAddEndPuncttrue
\mciteSetBstMidEndSepPunct{\mcitedefaultmidpunct}
{\mcitedefaultendpunct}{\mcitedefaultseppunct}\relax
\EndOfBibitem
\bibitem[Orús(2014)]{Orus:AnnPhys:2014}
Orús,~R. A practical introduction to tensor networks: Matrix product states
  and projected entangled pair states. \emph{Ann. Phys.} \textbf{2014},
  \emph{349}, 117--158\relax
\mciteBstWouldAddEndPuncttrue
\mciteSetBstMidEndSepPunct{\mcitedefaultmidpunct}
{\mcitedefaultendpunct}{\mcitedefaultseppunct}\relax
\EndOfBibitem
\bibitem[Makri(2020)]{a:Makri_20}
Makri,~N. Small matrix disentanglement of the path integral: Overcoming the
  exponential tensor scaling with memory length. \emph{J.~Chem.~Phys.}
  \textbf{2020}, \emph{152}, 041104\relax
\mciteBstWouldAddEndPuncttrue
\mciteSetBstMidEndSepPunct{\mcitedefaultmidpunct}
{\mcitedefaultendpunct}{\mcitedefaultseppunct}\relax
\EndOfBibitem
\bibitem[Richter and Fingerhut(2017)Richter, and Fingerhut]{a:Fingerhut_17}
Richter,~M.; Fingerhut,~B.~P. Coarse-grained representation of the quasi
  adiabatic propagator path integral for the treatment of non-Markovian
  long-time bath memory. \emph{J.~Chem.~Phys.} \textbf{2017}, \emph{146},
  214101\relax
\mciteBstWouldAddEndPuncttrue
\mciteSetBstMidEndSepPunct{\mcitedefaultmidpunct}
{\mcitedefaultendpunct}{\mcitedefaultseppunct}\relax
\EndOfBibitem
\bibitem[Richter and Fingerhut(2019)Richter, and Fingerhut]{a:Fingerhut_19}
Richter,~M.; Fingerhut,~B.~P. Coupled excitation energy and charge transfer
  dynamics in reaction centre inspired model systems. \emph{Faraday~Discuss.}
  \textbf{2019}, \emph{216}, 72\relax
\mciteBstWouldAddEndPuncttrue
\mciteSetBstMidEndSepPunct{\mcitedefaultmidpunct}
{\mcitedefaultendpunct}{\mcitedefaultseppunct}\relax
\EndOfBibitem
\bibitem[Sim and Makri(1997)Sim, and Makri]{a:Makri_97_II}
Sim,~E.; Makri,~N. Filtered propagator functional for iterative dynamics of
  quantum dissipative systems. \emph{Comput.~Phys.~Commun.} \textbf{1997},
  \emph{99}, 335\relax
\mciteBstWouldAddEndPuncttrue
\mciteSetBstMidEndSepPunct{\mcitedefaultmidpunct}
{\mcitedefaultendpunct}{\mcitedefaultseppunct}\relax
\EndOfBibitem
\bibitem[Sato(2019)]{a:Sato_19}
Sato,~Y. A scalable algorithm of numerical real-time path integral for quantum
  dissipative systems. \emph{J.~Chem.~Phys.} \textbf{2019}, \emph{150},
  224108\relax
\mciteBstWouldAddEndPuncttrue
\mciteSetBstMidEndSepPunct{\mcitedefaultmidpunct}
{\mcitedefaultendpunct}{\mcitedefaultseppunct}\relax
\EndOfBibitem
\bibitem[Topaler and Makri(1993)Topaler, and Makri]{a:Makri_93}
Topaler,~M.; Makri,~N. System-specific discrete variable representations for
  path integral calculations with quasi-adiabatic propagators.
  \emph{Chem.~Phys.~Lett.} \textbf{1993}, \emph{210}, 448\relax
\mciteBstWouldAddEndPuncttrue
\mciteSetBstMidEndSepPunct{\mcitedefaultmidpunct}
{\mcitedefaultendpunct}{\mcitedefaultseppunct}\relax
\EndOfBibitem
\bibitem[Caldeira and Leggett(1983)Caldeira, and Leggett]{a:Leggett_83}
Caldeira,~A.~O.; Leggett,~A.~J. Path integral approach to quantum Brownian
  motion. \emph{Physica~A} \textbf{1983}, \emph{121}, 587\relax
\mciteBstWouldAddEndPuncttrue
\mciteSetBstMidEndSepPunct{\mcitedefaultmidpunct}
{\mcitedefaultendpunct}{\mcitedefaultseppunct}\relax
\EndOfBibitem
\bibitem[Leggett \latin{et~al.}(1987)Leggett, Chakravarty, Dorsey, Fisher,
  Garg, and Zwerger]{a:Zwerger_87}
Leggett,~A.~J.; Chakravarty,~S.; Dorsey,~A.~T.; Fisher,~M.~P.~A.; Garg,~A.;
  Zwerger,~M. Dynamics of the dissipative two-state system.
  \emph{Rev.~Mod.~Phys.} \textbf{1987}, \emph{59}, 1\relax
\mciteBstWouldAddEndPuncttrue
\mciteSetBstMidEndSepPunct{\mcitedefaultmidpunct}
{\mcitedefaultendpunct}{\mcitedefaultseppunct}\relax
\EndOfBibitem
\bibitem[Makri(1992)]{a:Makri_92}
Makri,~N. Improved Feynman propagators on a grid and non-adiabatic corrections
  within the path integral framework. \emph{Chem.~Phys.~Lett.} \textbf{1992},
  \emph{193}, 435\relax
\mciteBstWouldAddEndPuncttrue
\mciteSetBstMidEndSepPunct{\mcitedefaultmidpunct}
{\mcitedefaultendpunct}{\mcitedefaultseppunct}\relax
\EndOfBibitem
\bibitem[Makri(1995)]{a:Makri_95}
Makri,~N. Numerical path integral techniques for long-time quantum dynamics of
  quantum dissipative systems. \emph{J.~Math.~Phys.} \textbf{1995}, \emph{36},
  2430\relax
\mciteBstWouldAddEndPuncttrue
\mciteSetBstMidEndSepPunct{\mcitedefaultmidpunct}
{\mcitedefaultendpunct}{\mcitedefaultseppunct}\relax
\EndOfBibitem
\bibitem[Sim and Makri(1997)Sim, and Makri]{a:Makri_97}
Sim,~E.; Makri,~N. Path integral simulation of charge transfer dynamics in
  photosynthetic reaction centers. \emph{J.~Phys.~Chem.} \textbf{1997},
  \emph{101}, 5446\relax
\mciteBstWouldAddEndPuncttrue
\mciteSetBstMidEndSepPunct{\mcitedefaultmidpunct}
{\mcitedefaultendpunct}{\mcitedefaultseppunct}\relax
\EndOfBibitem
\bibitem[Makarov and Makri(1994)Makarov, and Makri]{a:Makri_94}
Makarov,~D.~E.; Makri,~N. Path integrals for dissipative systems by tensor
  multiplication. Condensed phase quantum dynamics for arbitrarily long time.
  \emph{Chem.~Phys.~Lett.} \textbf{1994}, \emph{221}, 482\relax
\mciteBstWouldAddEndPuncttrue
\mciteSetBstMidEndSepPunct{\mcitedefaultmidpunct}
{\mcitedefaultendpunct}{\mcitedefaultseppunct}\relax
\EndOfBibitem
\bibitem[Makri(2012)]{a:Makri_12}
Makri,~N. Path integral renormalization for quantum dissipative dynamics with
  multiple timescales. \emph{Mol.~Phys.} \textbf{2012}, \emph{110}, 1001\relax
\mciteBstWouldAddEndPuncttrue
\mciteSetBstMidEndSepPunct{\mcitedefaultmidpunct}
{\mcitedefaultendpunct}{\mcitedefaultseppunct}\relax
\EndOfBibitem
\bibitem[Sim(2001)]{Sim:JCP:2001}
Sim,~E. Quantum dynamics for a system coupled to slow baths: On-the-fly
  filtered propagator method. \emph{J. Chem. Phys.} \textbf{2001}, \emph{115},
  4450--4456\relax
\mciteBstWouldAddEndPuncttrue
\mciteSetBstMidEndSepPunct{\mcitedefaultmidpunct}
{\mcitedefaultendpunct}{\mcitedefaultseppunct}\relax
\EndOfBibitem
\bibitem[wp:()]{wp:intelTBB}
\url{https://software.intel.com/content/www/us/en/develop/tools/threading-building-blocks.html}\relax
\mciteBstWouldAddEndPuncttrue
\mciteSetBstMidEndSepPunct{\mcitedefaultmidpunct}
{\mcitedefaultendpunct}{\mcitedefaultseppunct}\relax
\EndOfBibitem
\bibitem[Chin \latin{et~al.}(2013)Chin, Prior, Rosenbach, Caycedo-Soler,
  Huelga, and Plenio]{Chin:NatPhys:2013}
Chin,~A.~W.; Prior,~J.; Rosenbach,~R.; Caycedo-Soler,~F.; Huelga,~S.~F.;
  Plenio,~M.~B. {The role of non-equilibrium vibrational structures in
  electronic coherence and recoherence in pigment–protein complexes}.
  \emph{Nat. Phys.} \textbf{2013}, \emph{9}, 113--118\relax
\mciteBstWouldAddEndPuncttrue
\mciteSetBstMidEndSepPunct{\mcitedefaultmidpunct}
{\mcitedefaultendpunct}{\mcitedefaultseppunct}\relax
\EndOfBibitem
\bibitem[Thorwart \latin{et~al.}(2004)Thorwart, Paladino, and
  Grifoni]{a:Grifoni_04}
Thorwart,~M.; Paladino,~E.; Grifoni,~M. Dynamics of the spin-boson model with a
  structured environment. \emph{Chem.~Phys.} \textbf{2004}, \emph{296},
  333\relax
\mciteBstWouldAddEndPuncttrue
\mciteSetBstMidEndSepPunct{\mcitedefaultmidpunct}
{\mcitedefaultendpunct}{\mcitedefaultseppunct}\relax
\EndOfBibitem
\bibitem[Garg \latin{et~al.}(1985)Garg, Onuchic, and
  Ambegaokar]{a:Ambegaokar_1985}
Garg,~A.; Onuchic,~J.~N.; Ambegaokar,~V. Effect of friction on electron
  transfer in biomolecules. \emph{J.~Chem.~Phys.} \textbf{1985}, \emph{83},
  4491\relax
\mciteBstWouldAddEndPuncttrue
\mciteSetBstMidEndSepPunct{\mcitedefaultmidpunct}
{\mcitedefaultendpunct}{\mcitedefaultseppunct}\relax
\EndOfBibitem
\bibitem[Thorwart \latin{et~al.}(2000)Thorwart, Hartmann, Goychuk, and
  H\"anggi]{a:Haenggi_00}
Thorwart,~M.; Hartmann,~L.; Goychuk,~I.; H\"anggi,~P. Controlling decoherence
  of a two-level-atom in a lossy cavity. \emph{J.~Mod.~Opt.} \textbf{2000},
  \emph{47}, 2905\relax
\mciteBstWouldAddEndPuncttrue
\mciteSetBstMidEndSepPunct{\mcitedefaultmidpunct}
{\mcitedefaultendpunct}{\mcitedefaultseppunct}\relax
\EndOfBibitem
\bibitem[Iles-Smith \latin{et~al.}(2014)Iles-Smith, Lambert, and
  Nazir]{Iles-Smith:PhysRevA:2014}
Iles-Smith,~J.; Lambert,~N.; Nazir,~A. Environmental dynamics, correlations,
  and the emergence of noncanonical equilibrium states in open quantum systems.
  \emph{Phys. Rev. A} \textbf{2014}, \emph{90}, 032114\relax
\mciteBstWouldAddEndPuncttrue
\mciteSetBstMidEndSepPunct{\mcitedefaultmidpunct}
{\mcitedefaultendpunct}{\mcitedefaultseppunct}\relax
\EndOfBibitem
\end{mcitethebibliography}

\end{document}